\documentclass[pra,aps,nopacs,onecolumn,twoside,superscriptaddress]{revtex4}

\usepackage{amsmath,amsfonts,amssymb,caption,color,epsfig,graphics,graphicx,latexsym,mathrsfs,revsymb,theorem,url,verbatim,epstopdf,float}

\usepackage{hyperref}

\usepackage{amsmath,amsfonts,amssymb,caption,hyperref,color,epsfig,graphics,graphicx,latexsym,mathrsfs,revsymb,theorem,url,verbatim,epstopdf,cleveref}
\hypersetup{colorlinks,linkcolor={blue},citecolor={blue},urlcolor={red}}

\newtheorem{definition}{Definition}
\newtheorem{proposition}[definition]{Proposition}
\newtheorem{lemma}[definition]{Lemma}

\newtheorem{theorem}[definition]{Theorem}
\newtheorem{corollary}[definition]{Corollary}
\newtheorem{conjecture}[definition]{Conjecture}

\newtheorem{remark}[definition]{Remark}
\newtheorem{example}[definition]{Example}
\newtheorem{question}[definition]{Question}

\def\squareforqed{\hbox{\rlap{$\sqcap$}$\sqcup$}}
\def\qed{\ifmmode\squareforqed\else{\unskip\nobreak\hfil
\penalty50\hskip1em\null\nobreak\hfil\squareforqed
\parfillskip=0pt\finalhyphendemerits=0\endgraf}\fi}
\def\endenv{\ifmmode\;\else{\unskip\nobreak\hfil
\penalty50\hskip1em\null\nobreak\hfil\;
\parfillskip=0pt\finalhyphendemerits=0\endgraf}\fi}
\newenvironment{proof}{\noindent \textbf{{Proof.~} }}{\qed}
\def\Dbar{\leavevmode\lower.6ex\hbox to 0pt
{\hskip-.23ex\accent"16\hss}D}
\makeatletter
\def\url@leostyle{%
  \@ifundefined{selectfont}{\def\UrlFont{\sf}}{\def\UrlFont{\small\ttfamily}}}
\makeatother
\def\bcj{\begin{conjecture}}
\def\ecj{\end{conjecture}}
\def\bcr{\begin{corollary}}
\def\ecr{\end{corollary}}
\def\bd{\begin{definition}}
\def\ed{\end{definition}}
\def\bea{\begin{eqnarray}}
\def\eea{\end{eqnarray}}
\def\bem{\begin{enumerate}}
\def\eem{\end{enumerate}}
\def\bex{\begin{example}}
\def\eex{\end{example}}
\def\bim{\begin{itemize}}
\def\eim{\end{itemize}}
\def\bl{\begin{lemma}}
\def\el{\end{lemma}}
\def\bma{\begin{bmatrix}}
\def\ema{\end{bmatrix}}
\def\bpf{\begin{proof}}
\def\epf{\end{proof}}
\def\bpp{\begin{proposition}}
\def\epp{\end{proposition}}
\def\bqu{\begin{question}}
\def\equ{\end{question}}
\def\br{\begin{remark}}
\def\er{\end{remark}}
\def\bt{\begin{theorem}}
\def\et{\end{theorem}}

\def\btb{\begin{tabular}}
\def\etb{\end{tabular}}

\newcommand{\nc}{\newcommand}

\def\a{\alpha}
\def\b{\beta}

\def\e{\epsilon}

\def\i{\iota}

\def\r{\rho}
\def\s{\sigma}

\def\ps{\psi}

\def\G{\Gamma}

 \nc{\bbA}{\mathbb{A}} \nc{\bbB}{\mathbb{B}} \nc{\bbC}{\mathbb{C}}
 \nc{\bbD}{\mathbb{D}} \nc{\bbE}{\mathbb{E}} \nc{\bbF}{\mathbb{F}}
 \nc{\bbG}{\mathbb{G}} \nc{\bbH}{\mathbb{H}} \nc{\bbI}{\mathbb{I}}
 \nc{\bbJ}{\mathbb{J}} \nc{\bbK}{\mathbb{K}} \nc{\bbL}{\mathbb{L}}
 \nc{\bbM}{\mathbb{M}} \nc{\bbN}{\mathbb{N}} \nc{\bbO}{\mathbb{O}}
 \nc{\bbP}{\mathbb{P}} \nc{\bbQ}{\mathbb{Q}} \nc{\bbR}{\mathbb{R}}
 \nc{\bbS}{\mathbb{S}} \nc{\bbT}{\mathbb{T}} \nc{\bbU}{\mathbb{U}}
 \nc{\bbV}{\mathbb{V}} \nc{\bbW}{\mathbb{W}} \nc{\bbX}{\mathbb{X}}
 \nc{\bbZ}{\mathbb{Z}}

 \nc{\bA}{{\bf A}} \nc{\bB}{{\bf B}} \nc{\bC}{{\bf C}}
 \nc{\bD}{{\bf D}} \nc{\bE}{{\bf E}} \nc{\bF}{{\bf F}}
 \nc{\bG}{{\bf G}} \nc{\bH}{{\bf H}} \nc{\bI}{{\bf I}}
 \nc{\bJ}{{\bf J}} \nc{\bK}{{\bf K}} \nc{\bL}{{\bf L}}
 \nc{\bM}{{\bf M}} \nc{\bN}{{\bf N}} \nc{\bO}{{\bf O}}
 \nc{\bP}{{\bf P}} \nc{\bQ}{{\bf Q}} \nc{\bR}{{\bf R}}
 \nc{\bS}{{\bf S}} \nc{\bT}{{\bf T}} \nc{\bU}{{\bf U}}
 \nc{\bV}{{\bf V}} \nc{\bW}{{\bf W}} \nc{\bX}{{\bf X}}
 \nc{\bZ}{{\bf Z}}

\nc{\cA}{{\cal A}} \nc{\cB}{{\cal B}} \nc{\cC}{{\cal C}}
\nc{\cD}{{\cal D}} \nc{\cE}{{\cal E}} \nc{\cF}{{\cal F}}
\nc{\cG}{{\cal G}} \nc{\cH}{{\cal H}} \nc{\cI}{{\cal I}}
\nc{\cJ}{{\cal J}} \nc{\cK}{{\cal K}} \nc{\cL}{{\cal L}}
\nc{\cM}{{\cal M}} \nc{\cN}{{\cal N}} \nc{\cO}{{\cal O}}
\nc{\cP}{{\cal P}} \nc{\cQ}{{\cal Q}} \nc{\cR}{{\cal R}}
\nc{\cS}{{\cal S}} \nc{\cT}{{\cal T}} \nc{\cU}{{\cal U}}
\nc{\cV}{{\cal V}} \nc{\cW}{{\cal W}} \nc{\cX}{{\cal X}}
\nc{\cZ}{{\cal Z}}

\nc{\hA}{{\hat{A}}} \nc{\hB}{{\hat{B}}} \nc{\hC}{{\hat{C}}}
\nc{\hD}{{\hat{D}}} \nc{\hE}{{\hat{E}}} \nc{\hF}{{\hat{F}}}
\nc{\hG}{{\hat{G}}} \nc{\hH}{{\hat{H}}} \nc{\hI}{{\hat{I}}}
\nc{\hJ}{{\hat{J}}} \nc{\hK}{{\hat{K}}} \nc{\hL}{{\hat{L}}}
\nc{\hM}{{\hat{M}}} \nc{\hN}{{\hat{N}}} \nc{\hO}{{\hat{O}}}
\nc{\hP}{{\hat{P}}} \nc{\hR}{{\hat{R}}} \nc{\hS}{{\hat{S}}}
\nc{\hT}{{\hat{T}}} \nc{\hU}{{\hat{U}}} \nc{\hV}{{\hat{V}}}
\nc{\hW}{{\hat{W}}} \nc{\hX}{{\hat{X}}} \nc{\hZ}{{\hat{Z}}}

\nc{\hn}{{\hat{n}}}

\def\rank{\mathop{\rm rank}}

\def\ine{\mathop{\rm In}}
\def\ime{\mathop{\rm Im}}
\def\kere{\mathop{\rm Ker}}

\def\diag{\mathop{\rm diag}}
\def\dim{\mathop{\rm Dim}}

\def\min{\mathop{\rm min}}

\def\rank{\mathop{\rm rank}}

\def\dg{\dagger}

\def\ox{\otimes}

\newcommand{\bra}[1]{\langle#1|}
\newcommand{\ket}[1]{|#1\rangle}
\newcommand{\proj}[1]{| #1\rangle\!\langle #1 |}
\newcommand{\ketbra}[2]{|#1\rangle\!\langle#2|}

\newcommand{\abs}[1]{|#1|}

\newcommand{\tbc}{\red{TO BE CONTINUED...}}

\newcommand{\opp}{\red{OPEN PROBLEMS}.~}

\newcommand{\red}{\textcolor{red}}

\def\Dbar{\leavevmode\lower.6ex\hbox to 0pt
{\hskip-.23ex\accent"16\hss}D}

\begin{document}
\title{Inertia of partial transpose of positive semidefinite matrices}

\author{Yixuan Liang}\email[]{yixuanliang@buaa.edu.cn}
\affiliation{LMIB and School of Mathematical Sciences, Beihang University, Beijing 100191, China}

\author{Jiahao Yan}\email[]{21377242@buaa.edu.cn(corresponding author)}
\affiliation{LMIB and School of Mathematical Sciences, Beihang University, Beijing 100191, China}

\author{Dongran Si}\email[]{ terra@buaa.edu.cn(corresponding author)}
\affiliation{LMIB and School of Mathematical Sciences, Beihang University, Beijing 100191, China}

\author{Lin Chen}\email[]{linchen@buaa.edu.cn(corresponding author)}
\affiliation{LMIB and School of Mathematical Sciences, Beihang University, Beijing 100191, China}
\affiliation{International Research Institute for Multidisciplinary Science, Beihang University, Beijing 100191, China}

\date{\today}

\normalsize


\begin{abstract}
We show that the partial transpose of $9\times9$ positive semidefinite matrices do not have inertia $(4,1,4)$ and $(3,2,4)$. It solves an open problem in "LINEAR AND MULTILINEAR ALGEBRA. Changchun Feng et al, 2022". We apply our results to construct some inertia, as well as present the list of all possible inertia of partial transpose of $12\times12$ positive semidefinite matrices. 
\end{abstract}

\maketitle

Keywords: inertia, positive semidefinite matrix, non-positive-partial-transpose, NPT



\section{Introduction}

\label{sec:int}

The partial transpose of a positive semidefinite matrix, i.e., non-normalized quantum states has been useful for various fundamental tasks in quantum information theory, such as the separability problem, entanglement distillation \cite{Micha1998Mixed}, the construction of entanglement witnesses (EWs) \cite{GUHNE20091}. In particular, the states whose partial transpose are not positive semidefinite play a key role in quantum information processing, because many of such states may be converted to pure entanglement useful for quantum computing, teleportation and key distribution \cite{2007Quantum}. Such states are called the non-positive-partial-transpose (NPT) states. Hence, it is a meaningful task to investigate the NPT states from a theoretical and practical point of view \cite{2013Negative,2013Non}. 

One of such theoretical tools is the matrix inertia. It is a well known notion describing the number of negative, zero and positive eigenvalues for Hermitian matrices. For example, the partial transpose of NPT states has at least one negative eigenvalue. Further, the partial transpose of separable states remain positive semidefinite \cite{Peres1996}. Such states are positive-partial-transpose (PPT) states \cite{2012Qubit}, which have no negative eigenvalue. One can thus infer that NPT states are entangled states, though the converse holds for only the bipartite systems in $\bbC^2\otimes\bbC^2$ and $\bbC^2\otimes\bbC^3$ \cite{horodecki1997}.
For implementing entangled states and other physical tasks such as Bell inequalities, EWs were introduced and widely constructed experimentally \cite{2000Entanglement,2019Design, 2020Measurement}. Theoretically, the partial transpose of an NPT state is known as an EW. The inertia of EWs is thus a key factor of characterizing EWs. The inertia is also related to the entanglement measure such as the negativity \cite{2002Computable},
and the inertia of partial transpose of $2\times n$ and $3\times3$ NPT states has been partially investigated in \cite{2008Universal,2020Inertias,3x3inertia2022changchun}. Due to the mathematical difficulties of characterizing $3\times3$ systems, the existence of two inertias $(4,1,4)$ and $(3,2,4)$ of two-qutrit states remains unknown in \cite{3x3inertia2022changchun}. In this paper, we develop novel techniques and exclude their existence. This is presented in Theorem \ref{total}. Combining with the results in \cite{3x3inertia2022changchun}, we can conclude the list of  inertia of partial transpose of all two-qutrit NPT states. Then we apply our results to study the inertia of partial transpose of $3\times4$ NPT states.

The rest of this paper is organized as follows. In Sec. \ref{sec:pre} we present the fundamental concepts and facts on inertias. Then we respectively exclude the existence of inertia $(4,1,4)$ and $(3,2,4)$ in Sec. \ref{sec:414} and \ref{sec:324}. We apply our technique and results to study the inertia of partial transpose of $12\times12$ positive semidefinite matrices in Sec. \ref{sec:app}. Finally we conclude in Sec. \ref{sec:con}.

\section{preliminaries}
\label{sec:pre}

In this section, we introduce the basic facts and definitions used in later sections. We refer to 
$\bbC^d$ as the $d$-dimensional Hilbert space, and $I_d$ as the $d\times d$ unit matrix. A bipartite density matrix (i.e., a quantum state) $\r$ is a positive semidefinite matrix on the space $\cH_A\otimes\cH_B$. The partial transpose map on system $A$ sends $\r=\sum_{i,j}\ketbra{i}{j}\otimes\r_{ij}$ to $\r^\G=\sum_{i,j}\ketbra{i}{j}\otimes\r_{ji}$ with block matrices $\r_{ij}$. 
The state $\r$ is called non-positive-partial-transpose (NPT) when $\r^\Gamma$ has at least one negative eigenvalue. Otherwise, $\r$ is PPT. If $\rank\r=1$ then $\r$ is a pure state and can be written as $\r=\proj{\ps}$. The pure state $\ket{\ps}=\ket{a}\otimes\ket{b}:=\ket{a,b}$ is a bipartite product state. If $\ket{\ps}$ is not a product state then it is entangled. Investigating the number of product states in a given space plays a key role in our paper. For this purpose we present two facts, respectively stated as \cite[Proposition $6$]{Chen2013} and \cite[Lemma 5]{2020Inertias}. They respectively work for an arbitrary dimensional subspace and a two-qutrit subspace, i.e., $\bbC^3\otimes\bbC^3$.
\begin{lemma}
	\label{le:(m-1)(n-1)+1}
	Suppose the subspace $\mathcal{V} \subseteq \bbC^m \otimes \bbC^n$. If $\dim(\mathcal{V}) >(m-1)(n-1)$ then $\mathcal{V}$ contains at least one product
	vector. Furthermore, if $\dim(\mathcal{V}) >(m-1)(n-1)+1$ then $\mathcal{V}$ has infinitely many product vectors. 
\end{lemma}

\begin{lemma} 
	\label{le:pro in ker}
	Suppose $M$ is a two-qutrit state on $\mathbb{C}^m \ox \mathbb{C}^n$. Let $A$ be the non-positive eigen-space of $M^\G$, i.e., the sum of negative and zero eigen-spaces
	of $M^\G$. Then the product vectors in $A$ all belong to the zero eigen-space of $M^\G$. 
	
	
	
	\end{lemma}

In the rest of this section, we introduce facts on inertia of matrices. We refer to $\ine M=(a,b,c)$ as the inertia of the Hermitian matrix $M$. Here $a:=i_{-}(M)$, $b$ and $c:=i_{+}(M)$ are the numbers of negative, zero and positive eigenvalues of $M$, respectively. We refer to $A^\dag$, $A^T$ and $A^*$ as the conjugate transposition, transposition and conjugation of the matrix $A$, respectively. For a complex $a$, $a^*$ denotes the conjugation of $a$. We shall frequently use Sylvester's result that two Hermitian matrices have the same inertia when they are congruent (i.e., equivalent) via an invertible matrix. We refer to $A\simeq B$ as the congruence of  matrices $A$ and $B$. If the matrix is a product matrix, then the two Hermitian matrices are locally equivalent. The following lemma describes the relationship of product matrix and partial transpose.

\begin{lemma}\label{product tran}
Suppose $M$ is a $mn$ $\times$ $mn$ Hermitian matrix. If $A,C \in \bbC^{m\times m}$ and $B,D \in \bbC^{n\times n}$, then $$( (A\otimes B)M(C\otimes D))^{^\G}=(C^T\otimes B)M^{^\G}(A^T\otimes D).$$

Specially, $$( (A\otimes B)M(A^{\dag} \otimes B^{\dag}))^{^\G}=(A^* \otimes B)M^{^\G}(A^T\otimes B^{\dag}).$$
\end{lemma}

Moreover, product vector and partial transpose have the following relationship.

\begin{lemma}\label{vector tran}
Suppose $\phi=\beta \otimes \alpha$ is a product vector in $\mathbb{R}^m \otimes \mathbb{C}^n$, where $\beta \in \mathbb{R}^m$ and $\alpha \in \mathbb{C}^n$. If $M\in \mathbb{C}^{mn\times mn}$ is a Hermitian matrix, then
\begin{itemize}
    \item[(i)] $\phi^\dag M \phi=\phi^\dag M^\Gamma \phi$.
    \item[(ii)] if $M\geq 0$ and $M^\Gamma \phi=0$, then $M\phi=0$.
\end{itemize}
\end{lemma}
\begin{proof}
(i) Suppose $\beta=(b_1,\cdots,b_m)$ where $b_i\in \mathbb{R}$ for $ \forall 1 \leq i \leq m$ and $M=[M_{ij}]$. Then $$\phi^\dag M \phi=\sum\limits_{1\leq i,j\leq m} (b_i\alpha)^\dag M_{ij}(b_j\alpha)=\left(\sum\limits_{i=j}+\sum\limits_{i>j}+\sum\limits_{i<j} \right)b_ib_j\alpha^\dag M_{ij}\alpha.$$
We have $$\sum\limits_{i<j}b_ib_j\alpha^\dag M_{ij}\alpha=\sum\limits_{j<i}b_ib_j\alpha^\dag M_{ji}\alpha=\sum\limits_{i>j}b_ib_j\alpha^\dag M_{ij}^\dag \alpha.$$
So $$\phi^\dag M \phi=\sum\limits_{i=j}b_ib_j\alpha^\dag M_{ij}\alpha+\sum\limits_{i>j}b_ib_j\alpha^\dag(M_{ij}+M_{ij}^\dag)\alpha.$$
Similarly, $$\phi^\dag M^\Gamma \phi=\sum\limits_{i=j}b_ib_j\alpha^\dag M_{ji}\alpha+\sum\limits_{i>j}b_ib_j\alpha^\dag(M_{ji}+M_{ji}^\dag)\alpha.$$
Since for any $i>j$, $M_{ij}+M_{ij}^\dag=M_{ji}+M_{ji}^\dag$, we have $\phi^\dag M \phi=\phi^\dag M^\Gamma \phi.$

(ii) If $M^\Gamma \phi=0$, by (i), we have $\phi^\dag M \phi=0$. By $M\geq 0$, we can suppose $M=P^\dag P$, where $P$ is a $mn \times mn$ matrix. So $(P\phi)^\dag(P\phi)=0\Rightarrow\ P\phi=0 \Rightarrow\ M\phi=0.$
\end{proof}

We point out that the partial transpose of an NPT state $\r$ is an entanglement witness (EW) in quantum information. A bipartite EW is a non-semidefinite positive Hermitian matrix $H$ such that $\bra{a,b}H\ket{a,b}\ge0$ for any product vector $\ket{a,b}$. Hence, one can straightforwardly show that $\r^\G$ is an EW. The inertia of EWs has been investigated in \cite{2020Inertias,3x3inertia2022changchun}. For example, it has been shown that the number of positive eigenvalues of an EW is at least three. The set $\cN_{m,n}$ denotes all possible inertias of partial transpose of $m\times n$ NPT states. $\cN_{2,n}$ has been constructed in \cite{2020Inertias}. That is the following lemma discovered in 
\cite{3x3inertia2022changchun}.

\begin{lemma}\label{le:N_23}
\ $\mathcal{N}_{2,3}$ has exactly $4$ arrays,
	$(1,2,3)$, $(1,1,4)$, $(1,0,5)$, $(2,0,4)$.
\end{lemma}

We also need the following result as a partial fact on $\cN_{3,3}$ discovered in 
\cite{3x3inertia2022changchun}.

\begin{lemma}
	\label{le:N_33}
	(i) $\mathcal{N}_{3,3}$ has at least $13$ arrays,
	$(1,0,8)$, $(1,1,7)$, $(1,2,6)$, $(1,3,5)$, $(1,4,4)$, $(1,5,3)$, $(2,0,7)$, $(2,1,6)$, $(2,2,5)$, $(2,3,4)$, $(3,0,6)$, $(3,1,5)$, $(4,0,5)$.
	
	(ii) It is unknown whether $\mathcal{N}_{3,3}$ has the following two arrays,
	$(3,2,4)$, $(4,1,4)$. 
	
	(iii) $13\leq |\mathcal{N}_{3,3}|\leq 15$,  $\mathcal{N}_{3,3}$ is a subset of above (i) and (ii)'s arrays.
	
	(iv) The inertia of a  two-qutrit NPT state $\rho$ is one of the $13$ arrays in (i), when $\ker \rho^{\Gamma}$ has (iv.a) $|0,0\rangle$ and $|0,1\rangle$, or (iv.b) $\ket{0,0}$, $\ket{1,1}$ and $(\ket{0}+\ket{1})(\ket{0}+\ket{1})$, or (iv.c) $\ket{0,0}, \ket{1,1}, \ket{2}(\ket{0}+\ket{1}), (a\ket{0}+b\ket{1}+\ket{2})(\ket{0}+f\ket{1})$ satisfying the conditions $a,b\neq 0$  and $f\neq 0,1$.

(v) There are at least $(n-1)(2n-1)$ inertias in $\mathcal{N}_{3,n}$, i.e.,
		
		$|\mathcal{N}_{3,n}|\geq (n-1)(2n-1) $, for $\forall n\geq 2$.	
		
		Furthermore, these  $(n-1)(2n-1)$ inertias in $\mathcal{N}_{3,n}$ are as follows.
		
		$(1,3n-4-j,j+3)$, $\forall 0\leq j\leq 3n-4$,
		
		$(2,3n-6-j,j+4)$, $\forall 0\leq j\leq 3n-6$,
		
		...
		
		$(n-1,n-j,n+1+j)$, $\forall 0\leq j\leq n$.
\qed		
	\end{lemma}

\section{Exclusion of inertia $(4,1,4)$}
\label{sec:414}

In this section, we firstly present the main result of this paper, namely the exclusion of $(4,1,4)\in\cN_{3,3}$. We begin by presenting two observations on inertias of Hermitian matrices. The first one is from \cite{David2003Inertia}. It describes the inertias of linear operation of matrices. 
\begin{lemma}
	\label{sumdif lem}
	If $A$ and $B$ are $n\times n$ Hermitian matrices, then $$i_x(A+B) \le i_x(A)+i_x(B),$$ where $x=+$ or $-$.
	\qed
\end{lemma}

\begin{corollary}
\label{le:c0<=b0}
Suppose $M\ge0$ and $N$ is Hermitian, and they are of the same order. If $\ine N=(b_0, b_1, b_2)$ and $\ine (M+N)=(c_0, c_1, c_2)$, then $c_0\le b_0$ and $c_2\ge b_2$.
\end{corollary}
\begin{proof}
By Lemma \ref{sumdif lem}, we have $i_-(M)+i_-(N)\geq i_-(M+N)$ and $i_+(M+N)+i_+(-M)\geq i_+(N)$. By $M\geq 0$, we have $i_-(M)=i_+(-M)=0$. So $i_-(N)\geq i_-(M+N)$ and $i_+(M+N)\geq i_+(N)$.
\end{proof}

The second observation describes the inertias of product of matrices.

\begin{lemma}
\label{le:projection}
If $M$ is a $n\times n$ Hermitian matrix, $P\in \mathbb{C}^{n\times n}$ and $X=PMP^\dg$, then $i_x (M)\geq i_x (X)$ where $x=+$ or $-$. 
\end{lemma}
\begin{proof}
We consider the singular value decomposition $P=UDV$, where $U$ and $V$ are both $n\times n$ unitary matrices. We have $D=\diag (d_1,\cdots ,d_k,0, \cdots, 0)
$ where $k\le n$ and $d_j>0$ for $1\leq j\leq k$. Using Sylvester's theorem, we know that the inertia of Hermitian $N=VMV^\dg$ and that of $M$ are the same. Similarly we have  
\begin{eqnarray}
\label{eq:ineX}
\ine X=
\ine PMP^\dg
=\ine DVMV^\dg D^\dg
=\ine D N D^\dg.	
\end{eqnarray}
Then the assertion is equivalent to the claim $i_\pm(N)\ge i_\pm (DND^\dg)$. 

Let $N=\bma N_{11} & N_{12} \\ N_{21} & N_{22} \ema$, where $N_{11}$ is a $k\times k $ Hermitian matrix.
We have
$
DND^\dg \simeq N_{11}\oplus O_{n-k},	
$
which is also Hermitian. So $i_- (N_{11})=i_- (DND^\dg)$. Let $W$ be a unitary matrix such that $WN_{11}W^\dg=D_1\oplus D_2\oplus D_3$ is a real diagonal matrix, where $D_1<0$, $D_2=0$ and $D_3>0$. Then one can find a $d\times d$ invertible matrix $F$ such that 
\begin{eqnarray}
F(W\oplus I_{n-k})N(W^\dg\oplus I_{n-k})F^\dg
=D_1\oplus G.
\end{eqnarray}
Because $F$ and $W\oplus I_{n-k}$ are both invertible, the inertia of $N$ is the same as that of the Hermitian matrix $D_1\oplus G$. So $i_- (N)\ge i_- (D_1)=i_- (N_{11})=i_- (DND^\dg)$.

One can similarly prove 
$i_+(N)\ge i_+(DND^\dg)$. We have completed the proof of claim below \eqref{eq:ineX}. So the assertion of this lemma holds.
\end{proof}

\begin{corollary}\label{sub lem}
Suppose $M$ is a $d\times d$ Hermitian matrix. For any $ 1 \le s \le d$, let $M_s$ be the order-$s$ principal submatrix of $M$. Then $i_x (M)\geq i_x (M_s)$ with $x=+$ or $-$.
\end{corollary}
\begin{proof}
Suppose $P=I_s\oplus O_{d-s}$. Then, $PMP^\dag=M_s$. By Lemma \ref{le:projection}, we have $i_{\pm}(M) \ge i_{\pm}(M_{s})$.
\end{proof}

The above lemma shows that, the numbers of positive and negative eigenvalues of an Hermitian matrix do not increase under the congruence transformation. Then we use linear combinations of order-$2$ matrices to construct Hermitian matrix by the following lemma.

\begin{lemma}\label{22 hermit}
Suppose $S,B_1,B_2\in \mathbb{C}^{2\times 2}$  where $S\ge0$ and $S\neq O$. Then, there exist complex numbers $x,y,z$ such that $xS+yB_1+zB_2$ is Hermitian where one of $y$ and $z$ is nonzero. 
\end{lemma}
\begin{proof}
Suppose $S=\begin{pmatrix}
    a & c \\
    c^* & b
\end{pmatrix}$ where $a,b\in \mathbb{R}$ and $c\in \mathbb{C}$. By $S\ne O$, we can assume $a>0$. Suppose $B_1=\begin{pmatrix}
    d & e \\
    f & g
\end{pmatrix}$ and $B_2=\begin{pmatrix}
    h & k \\
    l & m
\end{pmatrix}$. We have $$M=xS+yB_1+zB_2=\begin{pmatrix}
    xa+yd+zh & xc+ye+zk \\
    xc^*+yf+zl & xb+yg+zm
\end{pmatrix}.$$

So $M=M^\dagger$ is equivalent to
\begin{align}\label{111}
 \begin{cases}
xa+yd+zh, xb+yg+zm\in \mathbb{R}, \\    xc+ye+zk=x^*c+y^*f^*+z^*l^*.
\end{cases}
\end{align}
Suppose $c=c_1+ic_2$ where $c_1,c_2\in \mathbb{R}$. One can similarly express the complex numbers $d,e,f,g,h,k,l,m,x,y,z$.
Then, condition (\ref{111}) is equivalent to the equations
$$ \begin{cases}
    a(x_1+ix_2)+(y_1+iy_2)(d_1+id_2)+(z_1+iz_2)(h_1+ih_2)\in \mathbb{R}, \\
    b(x_1+ix_2)+(y_1+iy_2)(g_1+ig_2)+(z_1+iz_2)(m_1+im_2)\in \mathbb{R}, \\
    (x_1+ix_2)(c_1+ic_2)+(y_1+iy_2)(e_1+ie_2)+(z_1+iz_2)(k_1+ik_2)=\\
    (x_1-ix_2)(c_1+ic_2)+(y_1-iy_2)(f_1-if_2)+(z_1-iz_2)(l_1-il_2),
\end{cases}$$
\begin{align}\label{22her 2}
\Leftrightarrow \begin{pmatrix}
    a & d_2 & d_1 & h_2 & h_1 \\
    b & g_2 & g_1 & m_2 & m_1 \\
    2c_2 & f_1-e_1 & e_2-f_2 & l_1-k_1 & k_2-l_2 \\
    2c_1 & e_2+f_2 & e_1+f_1 & k_2+l_2 & k_1+l_1
\end{pmatrix}
\begin{pmatrix}
    x_2 \\
    y_1 \\
    y_2 \\
    z_1 \\
    z_2
\end{pmatrix}=0.
\end{align}
Note that the coefficient matrix of the equation system \eqref{22her 2} is $4\times 5$. So \eqref{22her 2} has a nonzero solution $(x_2, y_1, y_2, z_1, z_2)
^T$. When $y_1=y_2=z_1=z_2=0$, according to $a>0$, we have $x_2=0$, which is a contradiction. So the nonzero solution satisfies that $y_1,y_2,z_1,z_2$ are not all zeros.
Let $x=ix_2,\ y=y_1+iy_2,\ z=z_1+iz_2$. So one of $y$ and $z$ is nonzero. By \eqref{111} and \eqref{22her 2}, $xS+yB_1+zB_2$ is a Hermitian matrix.
\end{proof}

Next, under the assumption that the inertia $(4,1,4)$ exists, we transform the state to a simpler form by the following lemma.

\begin{lemma}\label{first tran}
Suppose $M$ is a two-qutrit state whose partial transpose has inertia $(4,1,4)$. If $M$ exists, then we can choose the first row and column of $M$ and $M^\Gamma$ as zero, and the inertia of $M$ as $(0,1,8)$. 
\end{lemma}
\begin{proof}
Because the inertia of $M^\Gamma$ is $(4,1,4)$, by Lemma \ref{le:(m-1)(n-1)+1}, we know that there exists a product vector in the negative and zero space of $M^\G$. By Lemma \ref{le:pro in ker}, the product vector belongs to the zero eigen-space of $M^\G$. So we suppose $M^\G \beta \otimes \alpha=0$, where $\beta,\alpha\in \mathbb{C}^3$. Because there exist reversible matrices $A,B\in \mathbb{C}^{3\times 3}$ such that $A\ket{0}=\beta$ and $B\ket{0}=\alpha$, where $\ket{0}$ denotes the vector $(1,0,0)^T$. Let $N:=(A\otimes B)^\dag M^\G (A\otimes B)\simeq M^\G$. We have $N\ket{0}\otimes \ket{0}=0$. By Lemma \ref{product tran}, we have $\widetilde{M}:=N^\G=(A^T\otimes 
 B^\dag) M (A^* \otimes B)\simeq M\geq 0$. By Lemma \ref{vector tran}, we have $\widetilde{M}\ket{0}\otimes \ket{0}=0$. Now we have $\widetilde{M}\ket{0}\otimes \ket{0}=0$ and $\widetilde{M}^\G \ket{0}\otimes \ket{0}=0$. So the first row and column of $\widetilde{M}$ and $\widetilde{M}^\Gamma$ are zero. So we replace $M$ with $\widetilde{M}$ and come to the conclusion.

 Because $M^\G$ has inertia $(4,1,4)$, we can choose a small enough $\e>0$ such that $M+\e(0\oplus I_8)$ is a two-qutrit state of rank eight and the inertia of $M^\G$ is still $(4,1,4)$. We have proven the assertion.
\end{proof}

Through the lemmas, we are now in a position to show the main result of this section. 
\begin{theorem}
\label{thm:414}
Suppose $M$ is a two-qutrit state whose partial transpose has inertia $(4,1,4)$.

(i) If $Q$ is a $9\times9$ matrix of rank at least six, then the matrix $Q M^\Gamma Q^\dg$ is not positive semidefinite.

(ii) If $P$ is a $2\times3$ matrix of rank two, then the projected state $(P\otimes I_3)M(P^\dg \otimes I_3)$ is NPT.

(iii) The state $M$ does not exist, i.e. $(4,1,4)\notin \mathcal{N}_{3,3}$. 
\end{theorem}
\begin{proof}
(i) We disprove the claim. Suppose there exists an order-nine rank-$r (\ge6)$ matrix $Q$ such that the matrix $Q M^\Gamma Q^\dg$ is positive semidefinite. We consider the SVD of the matrix $Q=U(D\oplus0)V$, where $D>0$ is an order-$r$ matrix. Let 
\begin{eqnarray}
\label{eq:VrhoGammaVdag}	
VM^\G V^\dg=\bma M_{00} & M_{01} \\
M_{01}^\dg & M_{11} 
\ema,
\end{eqnarray}
where $M_{00}$ is an order-$r$ matrix. Because $QM^\G Q^\dg\ge0$, we obtain that $M_{00}\ge0$. 
Because $V$ is unitary, the inertia of $M^\G$ and $VM^\G V^\dg$ are both $(4,1,4)$. It follows from Lemma \ref{le:projection} that $R=\rank M_{00}\le 4$. 
We can find an order-$r$ unitary matrix $W$ such that $WM_{00}W^\dg=0_{r-R}\oplus D_R$ where $D_R>0$ is an order-$R$ matrix. We have
\begin{eqnarray}
\label{eq:VrhoGammaVdag1}	
X:=(W\oplus I_{r-4}) VM^\G V^\dg (W^\dg \oplus I_{r-4})=
\bma 
0_{r-R} & 0 & M_{010} \\
0 & D_R & M_{011} \\
M_{010}^\dg & M_{011}^\dg & M_{11}
\\ 
\ema.
\end{eqnarray}
Evidently, the inertia of $X$ is the same as that of $M^\G$. So $M_{010}$ has rank at least $\min\{r-R,9-r\}-1$. It implies that the matrix $X$ has positive eigenvalues with the number at least $\min\{r-R,9-r\}-1+R$. If $r-R\le 9-r$ then the number is at least
$r-1\ge5$. It is a contradiction with the fact that $X$ has the inertia $(4,1,4)$.

On the other hand, if $r-R> 9-r$ then 
$r-R=9-r+1\le 3$ by $\ine M^\G=(4,1,4)$. So the number of positive eigenvalues of $X$ is at least $9-r-1+R\ge5$. It is a contradiction with the fact that $X$ has the inertia $(4,1,4)$. We have proven the assertion.

(ii) The assertion is a corollary of (i) by setting $Q=P^\dg\otimes I_3$.

(iii) We disprove the claim. Suppose there exists a $9\times 9$ semidefinite Hermitian matrix $M$ such that $\ine M^\Gamma=(4,1,4)$. According to Lemma \ref{first tran}, we can suppose that all elements of the first row and column of both $M$ and $M^\Gamma$ are zero, and $\ine M=(0,1,8)$.

So we can assume $M=\begin{pmatrix}
    S & B_1 & B_2 \\
    B_1^\dag & M_{11} & M_{12} \\
    B_2^\dag & M_{12}^\dag & M_{22}
\end{pmatrix}$, where the first row and column of $S$, $B_1$ and $B_2$ are zero.
According to Lemma \ref{22 hermit}, there exist $x,y,z$ such that $xA+yB_1+zB_2$ is Hermitian and one of $y$ and $z$ is nonzero.
Let $P=\begin{pmatrix}
    1 & 0 & 0 \\
    x^* & y^* & z^*
\end{pmatrix}$. Since one of $y$ and $z$ is nonzero, we have $\rank(P)=2$. Then 
\begin{align*}
N=(P\otimes I_3)M(P^\dag\otimes I_3)=\begin{pmatrix}
    S & xS+yB_1+zB_2 \\
    x^*S+y^*B_1^\dg+z^*B_2^\dg & *
\end{pmatrix}\geq 0.
\end{align*}

Since $xS+yB_1+zB_2$ is Hermitian, we have $N^\Gamma=N\geq 0$. However, according to (ii), $N^\Gamma$ is not semidefinite, which is a contradiction. We have proven the theorem.
\end{proof}

\section{Exclusion of inertia $(3,2,4)$}
\label{sec:324}

In this section, we exclude the existence of two-qutrit states whose partial transpose has the inertia $(3,2,4)$. We also begin by presenting two observations on inertias of Hermitian matrices. The first one is about a kind of matrices with a special form.

\begin{lemma}
	\label{cross lem}
	If $M = 
	\begin{pmatrix}
		O_n & B  \\
		B^{\dag} & O_m  
	\end{pmatrix}$
	is an order-$(m+n)$ matrix where $B \in \mathbb{C}^{n\times m}$ and we suppose $r=\rank B$, then $i_+(M)=i_-(M)=r$.
\end{lemma}
\begin{proof}
	By the singular value decomposition, we assume $D=UBV$, where $U$ and $V$ are unitary matrices of order-$n$ and order-$m$, respectively. Meanwhile, $D=
	\begin{pmatrix}
		D_r & O_{r \times (m-r)}  \\
		O_{(n-r) \times r} & O_{(n-r) \times (m-r)}  
	\end{pmatrix}$, where $D_r=\diag(\sigma_1,\dots,\sigma_r)$ and $\sigma_i >0$.
	Then $M \simeq\
	\begin{pmatrix}
		U & O  \\
		O & V^\dag  
	\end{pmatrix} M 
	\begin{pmatrix}
		U^\dag & O \  \\
		O & V  
	\end{pmatrix}=
	\begin{pmatrix}
		O_n & D  \\
		D & O_m  
	\end{pmatrix}$. 
	Since $\ine
	\begin{pmatrix}
		0 & \sigma_i  \\
		\sigma_i & 0  
	\end{pmatrix} = (1,0,1)$, we have $\ine
	\begin{pmatrix}
		O_n & D  \\
		D & O_m  
	\end{pmatrix} = (r,m+n-2r,r)$. So $i_+(M)=i_-(M)=r$.
\end{proof}

The second observation is to judge the equality of inertias by order principal minor determinants of matrices. 

\begin{lemma}
	\label{det lem}
	Suppose $A,B$ are order-$n$ Hermitian matrices. For any $1 \le s \le n$, let $A_s$ and $B_s$ be the order-$s$ principal submatrices of $A$ and $B$, respectively. If for any $1\leq s\leq n$, $|A_s|,|B_s| \neq 0$ and $|A_s|,|B_s|$ have the same sign, then $\ine A = \ine B$.
\end{lemma}
\begin{proof}
We use the induction. If $s=1$, then $|A_1|$ and $|B_1|$ are nonzero and have the same sign. So $\ine A_1=\ine B_1=(1,0,0)$ or $(0,0,1)$. We assume $\ine A_s=\ine B_s=(p,0,s-p)$ and prove $\ine A_{s+1}= \ine B_{s+1}$. 

Note that $A_s$ is the principal submatrix of $A_{s+1}$. According to Corollary \ref{sub lem}, we have $i_+(A_{s+1}) \ge s-p$ and $ i_-(A_{s+1}) \ge p$. So $\ine A_{s+1}$ is $(p+1,0,s-p)$ or $(p,0,s-p+1)$. Similarly, $\ine B_{s+1}$ is $(p+1,0,s-p)$ or $(p,0,s-p+1)$.
	If $\ine A_{s+1} = (p+1,0,s-p)$, then $|A_{s+1}|$ and ${(-1)}^{p+1}$ have the same sign. If $\ine A_{s+1} = (p,0,s-p+1)$, then $|A_{s+1}|$ and ${(-1)}^p$ have the same sign. By $|A_{s+1}|=|B_{s+1}|$, we have $\ine A_{s+1} = \ine B_{s+1}$.
\end{proof}

Then, we give some lemmas about the inertia of $(3,2,4)$. We refer to $e_1,\cdots,e_n$ as the natural base of $\mathbb{C}^n$ throughout the rest of the paper.

\begin{lemma}\label{basic tran}
Suppose $M=[M_{ij}]\geq 0$ is a $3\times 3$ state and $\ine M^{\Gamma}=(3,2,4)$.
\begin{itemize}
    \item[(i)] We can make $Me_1=M^{\Gamma}e_1=0$.
    \item[(ii)] On the basis of (i), we can make that $M_{12}$ a Hermitian matrix.
    \item[(iii)] On the basis of (i) and (ii), we have $\rank(M_{11})=2$.
    \item[(iv)] On the basis of (i)--(iii), we can make $M_{11}=\diag(0,1,1)$, $M_{12}=\diag(0,\lambda_1,\lambda_2)$, where $\lambda_1,\lambda_2\in \mathbb{R}$.
    \item[(v)] On the basis of (i)--(iv), we can make $M_{13}=\begin{pmatrix}
        0 & & \\
         & 0 & a \\
         & b & u
    \end{pmatrix}$, where $a,b\in\mathbb{C}$ and $u\in\mathbb{R}$.
    \end{itemize}
\end{lemma}
\begin{proof}
    (i) Because the inertia of $M^\Gamma$ is $(3,2,4)$, we know that the negative and zero space of $M^\G$ has a product state. The proof is similar to the proof of Lemma \ref{first tran}.
    
    (ii) By $Me_1=M^{\Gamma}e_1=0$, the first row and column of $M_{11},M_{12},M_{13}$ are all zero. 
    If $M_{11}=O$, then by $M\geq 0$, we have $M_{12}=M_{13}=O$, which is a contradiction with $\ine M^\Gamma=(3,2,4)$.
    According to Lemma \ref{22 hermit}, there exist $x,y,z\in\mathbb{C}$ such that $xM_{11}+yM_{12}+zM_{13}$ is Hermitian and one of $y,z$ are nonzero. 
    
    We assume $y\neq 0$ and let $P=
    \begin{pmatrix}
        1 & x & 0 \\
        0 & y & 0 \\
        0 & z & 1 \\
    \end{pmatrix}$. Here $P$ is invertible. Let $N=(P\otimes I_3)^\dag M(P\otimes I_3)$. Then, $N_{12}=xM_{11}+yM_{12}+zM_{13}$ is a Hermitian matrix. By Lemma \ref{product tran}, we have $N\simeq M$ and $N^\G \simeq M^\G$. So $N\geq 0$ and $\ine N^\G=(3,2,4)$. We replace $M$ with $N$ and come to the conclusion.
    
    (iii) By (i), we have $\rank(M_{11})\le 2$. Suppose $M_{11}=(0)\oplus A$, where $A\in\mathbb{C}^{2\times 2}$. We prove $\rank M_{11} \neq 0,1$.
    
    (Case 1) If $\rank(M_{11})=0$, then $M_{11}=O$. Since $M\geq 0$, we have $M_{12}=M_{13}=O$, which is a contradiction with $\ine M^\Gamma=(3,2,4)$.
    
    (Case 2) If $\rank(M_{11})=1$, then there exists a unitary matrix $U\in \mathbb{C}^{2\times 2}$ such that $U^\dag AU=\diag(1,0)$. Let $P=I_1\oplus U$ and replace $M$ with $(I_3\otimes P)^\dag M(I_3\otimes P)$. So we have $M_{11}=\diag(0,1,0)$. By $M\ge 0$ and $M_{12}$ is Hermitian, $M$ has the form
    $$M=\left(\begin{array}{c|c|c}
         \begin{matrix}0 & \ & \ \\ \ & 1 & \ \\ \ & \ & 0 \end{matrix}&\
         \begin{matrix}0 & \ & \ \\ \ & p & \ \\ \ & \ & 0 \end{matrix}&\
         \begin{matrix}0 & \ & \ \\ \ & m & n \\ \ & \ & 0 \end{matrix}\\\hline
         \begin{matrix}0 & \ & \ \\ \ & p & \ \\ \ & \ & 0 \end{matrix}&\
         \begin{matrix}\ & \ & \ \\ \ & M_{22} & \ \\ \ & \ & \ \end{matrix}&\
         \begin{matrix}\ & \ & \ \\ \ & M_{23} & \ \\ \ & \ & \ \end{matrix}\\\hline
         \begin{matrix}0 & \ & \ \\ \ & m^* & \ \\ \ & n^* & 0 \end{matrix}&\
         \begin{matrix}\ & \ & \ \\ \ & M_{23}^\dag & \ \\ \ & \ & \ \end{matrix}&\
         \begin{matrix}\ & \ & \ \\ \ & M_{33} & \ \\ \ & \ & \ \end{matrix}
    \end{array}\right).$$
    \textbf{(Note)} Here, for convenience, we write the ordinary form and block form of the matrix in a formula. All blank spaces in the formula are zero. For the sake of brevity in the formulas, we omit them. Throughout the rest of the paper, we shall use the similar way to represent matrices.
    
    Let $Q=\begin{pmatrix}
        1 & -p & -m \\
         & 1 & \\
         & & 1
    \end{pmatrix}$ and replace $M$ with $(Q\otimes I_3)^\dag M(Q\otimes I_3)$. Then, $M$ and $M^\G$ have the form
    $$M=\left(\begin{array}{c|c|c}
         \begin{matrix}0 & \ & \ \\ \ & 1 & \ \\ \ & \ & 0 \end{matrix}&\
         \begin{matrix}0 & \ & \ \\ \ & 0 & \ \\ \ & \ & 0 \end{matrix}&\
         \begin{matrix}0 & \ & \ \\ \ & 0 & n \\ \ & \ & 0 \end{matrix}\\\hline
         \begin{matrix}0 & \ & \ \\ \ & 0 & \ \\ \ & \ & 0 \end{matrix}&\
         \begin{matrix}\ & \ & \ \\ \ & M_{22} & \ \\ \ & \ & \ \end{matrix}&\
         \begin{matrix}\ & \ & \ \\ \ & M_{23} & \ \\ \ & \ & \ \end{matrix}\\\hline
         \begin{matrix}0 & \ & \ \\ \ & 0 & \ \\ \ & n^* & 0 \end{matrix}&\
         \begin{matrix}\ & \ & \ \\ \ & M_{23}^\dag & \ \\ \ & \ & \ \end{matrix}&\
         \begin{matrix}\ & \ & \ \\ \ & M_{33} & \ \\ \ & \ & \ \end{matrix}
    \end{array}\right),
    M^{\Gamma}=\left(\begin{array}{c|c|c}
         \begin{matrix}0 & \ & \ \\ \ & 1 & \ \\ \ & \ & 0 \end{matrix}&\
         \begin{matrix}0 & \ & \ \\ \ & 0 & \ \\ \ & \ & 0 \end{matrix}&\
         \begin{matrix}0 & \ & \ \\ \ & 0 & \ \\ \ & n^* & 0 \end{matrix}\\\hline
         \begin{matrix}0 & \ & \ \\ \ & 0 & \ \\ \ & \ & 0 \end{matrix}&\
         \begin{matrix}\ & \ & \ \\ \ & M_{22} & \ \\ \ & \ & \ \end{matrix}&\
         \begin{matrix}\ & \ & \ \\ \ & M_{23}^\dag & \ \\ \ & \ & \ \end{matrix}\\\hline
         \begin{matrix}0 & \ & \ \\ \ & 0 & n \\ \ & \ & 0 \end{matrix}&\
         \begin{matrix}\ & \ & \ \\ \ & M_{23} & \ \\ \ & \ & \ \end{matrix}&\
         \begin{matrix}\ & \ & \ \\ \ & M_{33} & \ \\ \ & \ & \ \end{matrix}
    \end{array}\right).$$
    Let $\widetilde{M}=\left(\begin{array}{c|c|c}
         \begin{matrix}0 
         \end{matrix}&\
         \begin{matrix}0 & 0 & 0
         \end{matrix}&\
         \begin{matrix}0 & n^* & 0
         \end{matrix}\\\hline
         \begin{matrix}0 \\ 0 \\ 0
         \end{matrix}&\
         \begin{matrix}\ & \ & \ \\ \ & M_{22} & \ \\ \ & \ & \ \end{matrix}&\
         \begin{matrix}\ & \ & \ \\ \ & M_{23}^\dag & \ \\ \ & \ & \ \end{matrix}\\\hline
         \begin{matrix}0 \\ n \\ 0 
         \end{matrix}&\
         \begin{matrix}\ & \ & \ \\ \ & M_{23} & \ \\ \ & \ & \ \end{matrix}&\
         \begin{matrix}\ & \ & \ \\ \ & M_{33} & \ \\ \ & \ & \ \end{matrix}
    \end{array}\right)$ be an order-$7$ submatrix of $M^\G$. By $\ine M^{\Gamma}=(3,2,4)$, we have $\ine \widetilde{M}=(3,1,3)$.
    
    According to Lemma \ref{sumdif lem}, we have $i_-(\widetilde{M})\le i_-\left(\begin{matrix}
        0 & n^* \\
        n & 0
    \end{matrix}\right)+i_-\left(\begin{matrix}
        M_{22} & M_{23}^\dag \\
        M_{23} & M_{33}
    \end{matrix}\right)$. So $i_-\left(\begin{matrix}
        M_{22} & M_{23}^\dag \\
        M_{23} & M_{33} 
    \end{matrix}\right)\ge 2$. According to Lemma \ref{le:N_23}, $\ine \begin{pmatrix}
        M_{22} & M_{23}^\dag \\
        M_{23} & M_{33}
    \end{pmatrix}\in \mathcal{N}_{2,3}$. So $\ine \begin{pmatrix}
        M_{22} & M_{23}^\dag \\
        M_{23} & M_{33}
    \end{pmatrix}=(2,0,4)$. By Corollary \ref{sub lem}, we have $i_+(\widetilde{M})\geq i_+\left(\begin{matrix}
        M_{22} & M_{23}^\dag \\
        M_{23} & M_{33} 
    \end{matrix}\right)=4$. However, $i_+(\widetilde{M})=3<4$, which is a contradiction. 
    
    (iv) By (i), we can suppose $M_{11}=(0)\oplus A$ and $M_{12}=(0)\oplus B_1$. By (ii), $B_1$ is Hermitian. By (iii), we have $A>0$. So there exists a matrix $C\in\mathbb{C}^{2\times 2}$ such that $C^\dag AC=I_2$ and a unitary matrix $U\in\mathbb{C}^{2\times 2}$ such that $U^\dag C^\dag B_1CU=\diag(\lambda_1,\lambda_2)$. 
    Let $P=I_1\oplus CU$ and $N=(I_3\otimes P)^\dag M(I_3\otimes P)$. Then, we have $N_{11}=P^\dag M_{11}P=(0)\oplus (U^\dag C^\dag ACU)=\diag(0,1,1)$ and $N_{22}=P^\dag M_{12}P=(0)\oplus (U^\dag C^\dag B_1 CU)=\diag(0,\lambda_1,\lambda_2)$. Similarly to (ii), we replace $M$ with $N$ and come to the conclusion.
    
    (v) By (i), we can suppose $M_{13}=(0)\oplus B_2$ and $B_2=\begin{pmatrix}
        c & a \\
        b & d
    \end{pmatrix}$. 
    Let $P=\begin{pmatrix}
        1 & & -c \\
         & 1 & \\
         & & 1
    \end{pmatrix}$ and $N=(P\otimes I_3)^\dag M(P\otimes I_3)$.
    Then, $N_{13}=M_{33}-cM_{11}=\begin{pmatrix}
        0 & & \\
         & 0 & a \\
         & b & d-c
    \end{pmatrix}$. 
    
    If $d-c=0$, then $N$ has the form in (v). We replace $M$ with $N$ and come to the conclusion.
    
    If $d-c\neq 0$, then we suppose $Q=\begin{pmatrix}
        1 & & \\
         & 1 & \\
         & & \frac{1}{d-c}
    \end{pmatrix}$ and $\widetilde{N}=(I_3\otimes Q)^\dag N(I_3\otimes Q)$. Then, $\widetilde{N}_{13}=\frac{1}{d-c}N_{13}=\begin{pmatrix}
        0 & & \\
         & 0 & \frac{a}{d-c} \\
         & \frac{b}{d-c} & 1
    \end{pmatrix}$. We replace $M$ with $\widetilde{N}$ and come to the conclusion.
\end{proof}

The exclusion of inertia $(3,2,4)$ is much more complicated than that of $(4,1,4)$. So we need further simplification. Then we study the kernel space of a Hermitian matrix and its partial transpose by the following lemma and corollary.

\begin{lemma}\label{negetive change}
    Suppose $M\in \mathbb{C}^{n\times n}$ is a Hermitian matrix. For any $1\le s \le n$, suppose $A=(\alpha_1,\cdots,\alpha_s)\in \mathbb{C}^{n\times s}$, where $\alpha_l\in\mathbb{C}^n$ for $1\le l\le s$. Let $N=M-kAA^\dag$ where $k\in \mathbb{R}_+$, then 
    \begin{itemize}
        \item [(i)] if $\forall 1\le l\le s$, $\alpha_l\in \ime (M)$, then $\exists k_0>0$, such that $\forall 0<k<k_0$, $\ine N=\ine M$.
    \item[(ii)] if $\exists 1\le l \le s$, $\alpha_l\notin \ime (M)$, then $\forall k>0$, $i_-(N)>i_-(M)$.
       \end{itemize}
\end{lemma}

\begin{proof}
    Since $M$ is Hermitian, we have 
    \begin{align}\label{negtran 1}
    (\ime M)^{\perp}=\kere M,
    \end{align}
    where $(\ime M)^{\perp}$ denotes the orthogonal complement of $\ime M$. 
    There exists a unitary matrix $U=(\epsilon_1,\cdots,\epsilon_n)$ such that $U^\dag MU=D=\diag(\lambda_1,\cdots,\lambda_r,0,\cdots,0)$, where $\lambda_i\neq 0$. We suppose $\alpha_l=U\beta_l$ for any $1\le l\le s$. Then, we have $A=UB$ where $B=(\beta_1,\cdots,\beta_s)$. Thus, 
    \begin{align}\label{negtran 2}
    N=UDU^\dag-kUBB^\dag U^\dag \simeq D-kBB^\dag.
    \end{align}
    Note that $\kere M=\text{span}(\epsilon_{r+1},\cdots,\epsilon_n)$. By \eqref{negtran 1}, we have $\ime M=\text{span}(\epsilon_1,\cdots,\epsilon_r)$. 
    
    (i) If $\alpha_l\in \ime M$, by $\alpha_l=U\beta_l$, then the last $n-r$ element of $\beta_l$ are all zeros. So $B$ and $BB^\dag$ has the forms $$B=\begin{pmatrix}
        b_{11} & \cdots & b_{1r} \\
        \vdots & & \vdots \\
        b_{r1} & \cdots & b_{rr} \\
        0 & \cdots & 0 \\
        \vdots & & \vdots \\
        0 & \cdots & 0
    \end{pmatrix},\ BB^\dag=\begin{pmatrix}
        c_{11} & \cdots & c_{1r} & 0 & \cdots & 0 \\
        \vdots & & \vdots & \vdots & & \vdots \\
        c_{r1} & \cdots & c_{rr} & 0 & \cdots & 0\\
        0 & \cdots & 0 & 0 & \cdots & 0 \\
        \vdots & & \vdots & \vdots & & \vdots \\
        0 & \cdots & 0 & 0 & \cdots & 0
    \end{pmatrix}.$$
     By \eqref{negtran 2}, 
     \begin{align}\label{negtran 3}
     N\simeq D-kBB^\dag= \begin{pmatrix}
        \lambda_1-kc_{11} & \cdots & -kc_{1r} & 0 & \cdots & 0 \\
        \vdots & & \vdots & \vdots & & \vdots \\
        -kc_{r1} & \cdots & \lambda_r-kc_{rr} & 0 & \cdots & 0\\
        0 & \cdots & 0 & 0 & \cdots & 0 \\
        \vdots & & \vdots & \vdots & \ddots & \vdots \\
        0 & \cdots & 0 & 0 & \cdots & 0
    \end{pmatrix}.
    \end{align}
    Observing the form of $D-kBB^\dag$, there exists $k_0>0$ which is small enough such that for any $0<k\leq k_0$ all order principal minor determinants of $\begin{pmatrix}
        \lambda_1-kc_{11} & \cdots & -kc_{1r} \\
        \vdots & & \vdots  \\
        -kc_{r1} & \cdots & \lambda_r-kc_{rr}
    \end{pmatrix}$ and $\diag(\lambda_1,\cdots,\lambda_r)$ have the same sign. By Lemma \ref{det lem}, when $0<k\leq k_0$, $\ine D=\ine (D-kBB^\dag)\Rightarrow \ine M=\ine N$.
    
    (ii) If there exists $l_0$ such that $\alpha_{l_0}\notin \ime M$, by $\alpha_{l_0}=U\beta_{l_0}$, then the last $n-r$ elements of $\beta_{l_0}$ are not all zeros. We assume the $(r+1)$-th element of $\beta_{l_0}$ is not zero. Then $$BB^\dag=\begin{pmatrix}
        c_{11} & \cdots & c_{1r} & c_{1,r+1} & * & \cdots & *\\
        \vdots & & \vdots & \vdots & \vdots & & \vdots \\
        c_{r1} & \cdots & c_{rr} & c_{r,r+1} & * &\cdots & *\\
        c_{r+1,1} & \cdots & c_{r+1,r} & c_{r+1,r+1} & * & \cdots & * \\
        * & \cdots & * & * & * & \cdots & * \\
        \vdots & & \vdots & \vdots & \vdots & & \vdots \\
        * & \cdots & * & * & * & \cdots & *
    \end{pmatrix}, \ \text{where} \ c_{r+1,r+1}>0.$$ 
    We suppose that the order-$(r+1)$ principal submatrix of $D-kBB^\dag$ is $\widetilde{D}$. Then, $$\widetilde{D}= \begin{pmatrix}
        \lambda_1-kc_{11} & \cdots & -kc_{1r} & -kc_{1,r+1} \\
        \vdots & & \vdots & \vdots  \\
        -kc_{r1} & \cdots & \lambda_r-kc_{rr} & -kc_{r,r+1} \\
        -kc_{r+1,1} & \cdots & -kc_{r+1,r} & -kc_{r+1,r+1} 
    \end{pmatrix}.$$
    
    By symmetric elementary transformation, we can use $c_{r+1,r+1}$ to eliminate the elements of same row and column. So we have 
    $$\begin{pmatrix}
        c_{11} & \cdots & c_{1r} & c_{1,r+1} \\
        \vdots & & \vdots & \vdots \\
        c_{r1} & \cdots & c_{rr} & c_{r,r+1} \\
        c_{r+1,1} & \cdots & c_{r+1,r} & c_{r+1,r+1} 
    \end{pmatrix} \simeq \begin{pmatrix}
        d_{11} & \cdots & d_{1r} & 0 \\
        \vdots & & \vdots & \vdots \\
        d_{r1} & \cdots & d_{rr} & 0 \\
        0 & \cdots & 0 & c_{r+1,r+1}
    \end{pmatrix}.$$ 
    Thus, $$\widetilde{D} \simeq \begin{pmatrix}
        \lambda_1-kd_{11} & \cdots & -kd_{1r} & 0 \\
        \vdots & & \vdots & \vdots \\
        -kd_{r1} & \cdots & \lambda_r-kd_{rr} & 0 \\
        0 & \cdots & 0 & -kc_{r+1,r+1}
    \end{pmatrix}.$$ 
    Note that there exists $k_0>0$ which is small enough such that for any $0<k\leq k_0$, all order principal minor determinants of $\begin{pmatrix}
        \lambda_1-kd_{11} & \cdots & -kd_{1r} \\
        \vdots & & \vdots  \\
        -kd_{r1} & \cdots & \lambda_r-kd_{rr}
    \end{pmatrix}$ and $\diag(\lambda_1,\cdots,\lambda_r)$ have the same sign. By Lemma \ref{det lem},
     we have
    $$i_- \begin{pmatrix}
        \lambda_1-kd_{11} & \cdots & -kd_{1r} \\
        \vdots & & \vdots  \\
        -kd_{r1} & \cdots & \lambda_r-kd_{rr}\end{pmatrix}=i_-(\diag(\lambda_1,\cdots,\lambda_r))=i_-(D).$$
    So $$i_-(\widetilde{D})=i_- \begin{pmatrix}
        \lambda_1-kd_{11} & \cdots & -kd_{1r} \\
        \vdots & & \vdots  \\
        -kd_{r1} & \cdots & \lambda_r-kd_{rr}\end{pmatrix}+1=i_-(D)+1>i_-(D).$$
    By Corollary \eqref{sub lem}, we have $i_- (D-kBB^\dag)\geq i_-(\widetilde{D})$. So $i_-(N)>i_-(M)$.

For any $k> k_0$, since $NN^\dag\geq 0$, by Lemma \eqref{sumdif lem}, we have $i_-(D-kBB^\dag)>i_-(D-k_0BB^\dag)$. So $i_-(N)>i_-(M)$ holds for any $k>0$. 
\end{proof}

\begin{corollary}\label{ker trans}
    Suppose $M=[M_{ij}]\geq 0$ is a $3\times 3$ state. If $M_{12}$ is Hermitian and $\ine M^{\Gamma}=(3,2,4)$, then $\kere M^{\Gamma} \subseteq \kere M$. 
\end{corollary}

\begin{proof}
    There exists a unitary matrix $U=(\epsilon_1,\cdots,\epsilon_9)$ such that $U^\dag MU=D=\diag(\lambda_1,\cdots,\lambda_r,0,\cdots,0)$, where $\lambda_i>0$ for any $1\le i\le r$. Let $A=(\sqrt{\lambda_1}\epsilon_1,\cdots,\sqrt{\lambda_r}\epsilon_r)$, then $M=U DU^\dag=AA^\dag$. Note that $\kere M=\text{span}(\epsilon_{r+1},\cdots,\epsilon_{9})$. By $(\ime M)^{\perp}=\kere M$, we have $\ime M=\text{span}(\epsilon_1,\cdots,\epsilon_r)$. 
    
    Then we prove $\ime M\subseteq \ime M^{\Gamma}$ by contradiction. Assume that there exists $1\le i\le r$ such that $\epsilon_i\notin \ime M^{\Gamma}$. By Lemma \ref{negetive change}(ii), we have \begin{align}\label{kertran 1} i_-(M^{\Gamma}-AA^\dag)>i_-(M^{\Gamma})=3. \end{align} However, by $M_{12}=M_{12}^\dag$, we have $$M^{\Gamma}-M=\begin{pmatrix}
        0 & 0 & M_{13}^\dag-M_{13} \\
        0 & 0 & M_{23}^\dag-M_{23} \\
        M_{13}-M_{13}^\dag & M_{23}-M_{23}^\dag & 0
    \end{pmatrix}.$$ Note that $\rank \left(\begin{matrix}
        M_{13}^\dag-M_{13} \\
        M_{23}^\dag-M_{23}
    \end{matrix}\right)\le 3$. By Lemma \ref{cross lem}, we have $i_-(M^{\Gamma}-M)\le 3$, which is a contradiction with \eqref{kertran 1}. So $\ime M\subseteq \ime M^{\Gamma}\Rightarrow (\ime M)^{\perp}\supseteq (\ime M^{\Gamma})^{\perp}\Rightarrow \kere M^{\Gamma}\subseteq \kere M$.
\end{proof}

By Lemma \ref{basic tran} and Corollary \ref{ker trans}, we can determine the kernel space of the state $M$ in two cases. That is the following lemma.

\begin{lemma}\label{end trans}
 Suppose $M=[M_{ij}]\geq 0$ is a $3\times 3$ state. If $\ine M^{\Gamma}=(3,2,4)$, then we can choose $M$ to let one of the following cases holds:
 
     \item[(Case 1)] $Me_1=M^\Gamma e_1=Me_5=M^\Gamma e_5=0$.
     
    \item[(Case 2)] $Me_{1}=M^\Gamma e_1=M(e_2+e_6)=M^\Gamma(e_2+e_6)=0$.

\end{lemma}

\begin{proof}
    According to Lemma \ref{basic tran}(v), we can make 
    \begin{align}
    Me_1&=M^\Gamma e_1=0,\notag \\
    M_{11}&=\diag(0,1,1),\notag \\ M_{12}&=\diag(0,\lambda_1,\lambda_2) \ \text{and}\notag \\
    M_{13}&=\begin{pmatrix}
        0 & & \\
         & 0 & a \\
         & b & u
    \end{pmatrix}\label{endtran 0.3},
    \end{align}
    where $\lambda_1,\lambda_2,u\in \mathbb{R}$ and $a,b\in \mathbb{C}$. By $\ine M^{\Gamma}=(3,2,4)$, we suppose $\kere M^\Gamma=\text{span}(e_1,X)$, where $X\in \mathbb{C}^9$ is linearly independent with $e_1$. So $M^\Gamma X=0$. According to Corollary \ref{ker trans}, we have 
    \begin{align}
        &MX=0, \label{endtran 0.4} \\
        &(M^\Gamma-M)X=0.\label{endtran 0.5}
    \end{align}
    
    We suppose $X=(x_i)_{9\times 1}$ and try to prove that $x_7=x_8=x_9=0$.
    Since $M_{12}$ is Hermitian in \eqref{endtran 0.3}, we have 
    \begin{align}\label{endtran 0.9}
    M^\G-M=\begin{pmatrix}
        0 & 0 & M_{13}^\dag-M_{13} \\
        0 & 0 & M_{23}^\dag-M_{23} \\
        M_{13}-M_{13}^\dag & M_{23}-M_{23}^\dag & 0
    \end{pmatrix}.
    \end{align}
    By Lemma \ref{sumdif lem}, we have \begin{align}\label{endtran 1}
        i_-(M^\Gamma -M)\ge 3.
    \end{align} 
    By \eqref{endtran 0.5}, we have
    \begin{align}
        (M_{13}^\dag-M_{13})\begin{pmatrix}
        x_7 \\
        x_8 \\
        x_9\\
    \end{pmatrix}=0, \label{endtran 2} \\
     (M_{23}^\dag-M_{23})\begin{pmatrix}
        x_7 \\
        x_8 \\
        x_9
    \end{pmatrix}=0 \label{endtran 3}.
    \end{align}

    According to \eqref{endtran 0.3}, we classify to discuss.
    
    (i) If $a=b^*$, then $M_{13}=M_{13}^\dag$. By  \eqref{endtran 0.9}, \eqref{endtran 1} and Lemma \ref{cross lem}, we have $\rank\left(\begin{matrix}
        M_{13}^\dag-M_{13} \\
        M_{23}^\dag-M_{23}
    \end{matrix}\right)= 3$. So $\rank(M_{23}^\dag-M_{23})=3$. By \eqref{endtran 3}, $x_7=x_8=x_9=0$.
    
    (ii) If $a\neq b^*$, then we have $M_{13}^\dag-M_{13}=\begin{pmatrix}
        0 & & \\
         & 0 & b^*-a \\
         & a^*-b & 0
    \end{pmatrix}$. By \eqref{endtran 2}, we have $x_8=x_9=0$. Next, if the first column of $M_{23}^\dag-M_{23}$ are zeros, then $\rank\begin{pmatrix}
        M_{13}^\dag-M_{13} \\
        M_{23}^\dag-M_{23}
    \end{pmatrix}\leq 2$. By \eqref{endtran 0.9} and Lemma \ref{cross lem}, we have $i_-(M^\G-M)\leq 2$, which is a contradiction with \eqref{endtran 1}. So the first column of $M_{23}^\dag-M_{23}$ are not all zeros. Thus, $x_7=0$.
    
    Now we have $x_7=x_8=x_9=0$. So  $X=\ket{0}\otimes \alpha +\ket{1}\otimes \beta$ where $\alpha,\beta \in \mathbb{C}^3$. Here $\ket{0}$ and $\ket{1}$ denote the vectors $(1,0,0)^T$ and $(0,1,0)^T$.
    
    \textbf{(Case 1)} Suppose $\ket{0},\alpha,\beta$ are linearly dependent. First, we prove that $\ket{0}, \beta$ are linearly independent. If $\beta=k\ket{0}$, then $x_5=x_6=0$. In \eqref{endtran 0.4}, observing the second and third row of $M$, we have $$\begin{cases}
        x_2+\lambda_1 x_5=0 \\
        x_3+\lambda_2 x_6=0
    \end{cases}\Rightarrow \begin{cases}
        x_2=0 \\
        x_3=0
    \end{cases}.$$ 
    By \eqref{endtran 0.5} and \eqref{endtran 0.9}, we have $$(M_{23}-M_{23}^\dag)(x_4,x_5,x_6)^T=0.$$
    Because the first column of $M_{23}-M_{23}^\dag$ are not all zeros, we have $x_4=0$. Now, $x_i=0\ \text{for}\ 2 \leq i \leq 9$, which is a contradiction with the fact that $X,e_1$ are linearly independent.
    
    Now, we prove that $\ket{0}, \beta$ are linearly independent. So we can assume $\alpha=k_1\ket{0}+k_2\beta$. Then, $$X=k_1\ket{0}\otimes\ket{0}+k_2\ket{0}\otimes \beta+\ket{1}\otimes \beta.$$ Let $$Y=X-k_1e_1=(k_2\ket{0}+\ket{1})\otimes \beta.$$ 
    Then $Y\in \kere M^\Gamma \subseteq \kere M$. There exist invertible matrices $P,Q\in \mathbb{C}^{3\times 3}$ such that $P\ket{0}=\ket{0}$, $P\ket{1}=k_2\ket{0}+\ket{1}$, $Q\ket{0}=\ket{0}$ and $Q\ket{1}=\beta$. Let $N=(P\otimes Q)^\dag M^\Gamma(P\otimes Q)$. Then we have $$\begin{cases}
        N^\Gamma(\ket{0}\otimes \ket{0})=0 \\
        N^\Gamma(\ket{1}\otimes \ket{1})=0
    \end{cases}.$$ Then we replace $M$ with $N$. According to Lemma \ref{vector tran}, $\ket{0}\otimes \ket{0}$, $\ket{1}\otimes \ket{1}\in \kere M$. So we have $M^\Gamma e_1=M^\Gamma e_5=Me_1=Me_5=0$.\\
    \textbf{(Case 2)} Suppose $\ket{0}$,$\alpha$,$\beta$ are linearly independent. There exists an invertible matrix $Q\in\mathbb{C}^{3\times 3}$ such that $Q\ket{0}=\ket{0}$, $Q\ket{1}=\alpha$ and $Q\ket{2}=\beta$. We replace $M^\Gamma$ with $(I_3 \otimes Q)^\dag M^\Gamma (I_3\otimes Q)$. Similar to Case 1, we have $M^\Gamma e_1=M^\Gamma(e_2+e_6)=Me_1=M(e_2+e_6)=0$.
\end{proof}

According to the two cases in Lemma \ref{end trans}, we respectively exclude the existence of inertia $(3,2,4)$ by the following lemmas.

\begin{lemma}
	\label{lem:324 1}
  Suppose $M=[M_{ij}]\geq 0$ is a $3\times 3$ state and $\ine M^{\Gamma}=(3,2,4)$. If $Me_1 = M^{\Gamma}e_1 = Me_5= M^\Gamma e_5 = 0$, then
	\begin{itemize}
	    \item[(i)] we can make $\ine M=(0,2,7)$.
	
	\item[(ii)] on the basic of (i), we can make that $M$ has the form
 $$M=\left(\begin{array}{c|c|c}
		\begin{matrix}0 & & \\  &* & \\ & &* \end{matrix}&\ 
		\begin{matrix}0 & & \\  &0 & \\ & &* \end{matrix}&\ 
		\begin{matrix}0 & & \\  &* &* \\ &* &* \end{matrix}\\\hline 
		\begin{matrix}0 & & \\  &0 & \\ & &* \end{matrix}&
		\begin{matrix}* & & \\  &0 & \\ & &* \end{matrix}&\ 
		\begin{matrix}* & &* \\  &0 & \\ *& &* \end{matrix}\\\hline 
		\begin{matrix}0 & & \\  &* &* \\ &* &* \end{matrix}&\ 
		\begin{matrix}* & &* \\  &0 & \\ *& &* \end{matrix}&\ 
		\begin{matrix}* &* &* \\  *&* &* \\ *&* &* \end{matrix}
	\end{array}\right).$$
 \item[(iii)] on the basic of (i) and (ii), we can make that $M$ has the form
 $$M=\left(\begin{array}{c|c|c}
		\begin{matrix}0 &\  &\  \\  \ &* &\  \\ \ &\  &* \end{matrix}&\ 
		\begin{matrix}0 &\  &\  \\  \ &0 &\  \\ \ &\  &z \end{matrix}&\ 
		\begin{matrix}0 &\  &\  \\  \ &0 &* \\ \ &* &* \end{matrix}\\\hline 
		\begin{matrix}0 &\  &\  \\  \ &0 &\  \\ \ &\  &z \end{matrix}&
		\begin{matrix}* &\  &\  \\  \ &0 &\  \\ \ &\  &* \end{matrix}&\ 
		\begin{matrix}0 &\  &* \\  \ &0 &\  \\ *&\  &* \end{matrix}\\\hline 
		\begin{matrix}0 &\  &\  \\  \ &0 &* \\ \ &* &* \end{matrix}&\ 
		\begin{matrix}0 &\  &* \\  \ &0 &\  \\ *&\  &* \end{matrix}&\ 
		\begin{matrix}* &* &* \\  *&* &* \\ *&* &* \end{matrix}\end{array}\right),$$where $z\in \mathbb{R}$.
   \item[(iv)] there exists a contradiction. So $\ine M^\G \neq (3,2,4).$
 	\end{itemize}
\end{lemma}

\textbf{(Note)} In (ii) and (iii), we represent $M$ in a special way. The ``$*$'' in formulas denotes unknown elements, which can be either $0$ or not. All blank spaces denote zero. Throughout the rest of the paper, we shall use the similar way to represent matrices. 

\begin{proof}
	(i) By $Me_1 = M^{\Gamma}e_1=Me_5= M^\Gamma e_5 = 0$, we know that the first and fifth columns of $M$ and $M^\Gamma$ are zeros. Since $M$ is Hermitian, $M$ has the form
	\begin{align}\label{324 1}
	M = 
	\left(\begin{array}{c|c|c}
		\begin{matrix}0 & & \\  &* &* \\ &* &* \end{matrix}&\ 
		\begin{matrix}0 & & \\  &0 & \\ & &* \end{matrix}&\ 
		\begin{matrix}0 & & \\  &* &* \\ &* &* \end{matrix}\\\hline 
		\begin{matrix}0 & & \\  &0 & \\ & &* \end{matrix}&
		\begin{matrix}* & &* \\  &0 & \\ *& &* \end{matrix}&\ 
		\begin{matrix}* & &* \\  &0 & \\ *& &* \end{matrix}\\\hline 
		\begin{matrix}0 & & \\  &* &* \\ &* &* \end{matrix}&\ 
		\begin{matrix}* & &* \\  &0 & \\ *& &* \end{matrix}&\ 
		\begin{matrix}* &* &* \\  *&* &* \\ *&* &* \end{matrix}
	\end{array}\right).
	\end{align}

Assume $N=M + \delta \diag(0, 1, 1, 1, 0, 1, 1, 1, 1)$. By $\ine M^\Gamma = (3,2,4)$ and $M\geq 0$, when $\delta$ is small enough, we have $\ine N^\Gamma = (3,2,4)$ and $\ine N = (0,2,7)$. We replace $M$ with $N$. So we can make
      $\ine M = (0,2,7)$.
	
	(ii) Note that in \eqref{324 1},$$M_{11}=\begin{pmatrix}
	    0 &  &  \\
             & m_{22} & m_{23} \\
             & m_{23}^* & m_{33} 
	\end{pmatrix},\ 
 M_{22}=\begin{pmatrix}
	    m_{44} &  & m_{46} \\
             & 0 &  \\
            m_{46}^* &  & m_{66} 
	\end{pmatrix}.$$
 By $\ine M = (0,2,7)$, $\ine M_{11}=\ine M_{22}=(0,1,2)$. So we have $m_{22},m_{44}>0$. Let $P=\begin{pmatrix}
		1 & \ &-\frac{m_{46}}{m_{44}} \\
		  & 1 &-\frac{m_{23}}{m_{22}} \\
		  &   &1 \\  
	\end{pmatrix}$ and we have $P^\dag M_{11}P$ and $P^\dag M_{22}P$ are diagonal matrices. Moreover, $P^\dag M_{12}P$, $P^\dag M_{13}P$ and $P^\dag M_{23}P$ have the form $\begin{pmatrix}0 & & \\  &0 & \\ & &* \end{pmatrix}$, $\begin{pmatrix}0 & & \\  & * & * \\ & * & * \end{pmatrix}$ and $\begin{pmatrix}* & 
 &* \\  &0 & \\ * & &* \end{pmatrix}$, respectively. Let $N={(I_3 \otimes P)}^\dag M (I_3 \otimes P)$ and replace $M$ with $N$. We come to the conclusion.

	(iii) By (ii), we can make$$M = 
	\left(\begin{array}{c|c|c}
		\begin{matrix}0 &\  &\  \\  \ &m_{22} &\  \\ \ &\  &* \end{matrix}&\ 
		\begin{matrix}0 &\  &\  \\  \ &0 &\  \\ \ &\  &* \end{matrix}&\ 
		\begin{matrix}0 &\  &\  \\  \ &m_{28} &* \\ \ &* &* \end{matrix}\\\hline 
		\begin{matrix}0 &\  &\  \\  \ &0 &\  \\ \ &\  &* \end{matrix}&
		\begin{matrix}m_{44} &\  &\  \\  \ &0 &\  \\ \ &\  &* \end{matrix}&\ 
		\begin{matrix}m_{47} &\  &* \\  \ &0 &\  \\ *&\  &* \end{matrix}\\\hline 
		\begin{matrix}0 &\  &\  \\  \ &m_{28}^* &* \\ \ &* &* \end{matrix}&\ 
		\begin{matrix}m_{47}^* &\  &* \\  \ &0 &\  \\ *&\  &* \end{matrix}&\ 
		\begin{matrix}* &* &* \\  *&* &* \\ *&* &* \end{matrix}
	\end{array}\right),$$
 where $m_{22},m_{44}>0$.
 Let $Q = 
	\begin{pmatrix}
		1 & \ &-\frac{m_{28}}{m_{22}} \\
		& 1 &-\frac{m_{47}}{m_{44}} \\
		&   &1 \\  
	\end{pmatrix}$ and $N={(Q \otimes I_3)}^\dag M (Q \otimes I_3)$. Then, $N_{13}=M_{13}-\dfrac{m_{28}}{m_{22}}M_{11}-\dfrac{m_{47}}{m_{44}}M_{12}$ has the form $\begin{pmatrix}0 & & \\  &0 &* \\ &* &* \end{pmatrix}$. $N_{23}=M_{23}-\dfrac{m_{47}}{m_{44}}M_{22}-\dfrac{m_{28}}{m_{22}}M_{12}^*$ has the form $\begin{pmatrix}0 & &* \\  &0 & \\ * & &* \end{pmatrix}$. Moreover, we have $N_{11}=M_{11},N_{12}=M_{12}$ and $N_{22}=M_{22}$. So $N$ has the form
	\begin{align}	
	\label{324 4}
	N = 
	\left(\begin{array}{c|c|c}
		\begin{matrix}0 &\  &\  \\  \ &* &\  \\ \ &\  &* \end{matrix}&\ 
		\begin{matrix}0 &\  &\  \\  \ &0 &\  \\ \ &\  &t \end{matrix}&\ 
		\begin{matrix}0 &\  &\  \\  \ &0 &* \\ \ &* &* \end{matrix}\\\hline 
		\begin{matrix}0 &\  &\  \\  \ &0 &\  \\ \ &\  &t^* \end{matrix}&
		\begin{matrix}* &\  &\  \\  \ &0 &\  \\ \ &\  &* \end{matrix}&\ 
		\begin{matrix}0 &\  &* \\  \ &0 &\  \\ *&\  &* \end{matrix}\\\hline 
		\begin{matrix}0 &\  &\  \\  \ &0 &* \\ \ &* &* \end{matrix}&\ 
		\begin{matrix}0 &\  &* \\  \ &0 &\  \\ *&\  &* \end{matrix}&\ 
		\begin{matrix}* &* &* \\  *&* &* \\ *&* &* \end{matrix}
	\end{array}\right)
	\end{align}

	  If $t=0$ in \eqref{324 4}, $N$ has the form in (iii). We replace $M$ with $N$ and come to the conclusion. 
   
   If $t\neq 0$, then we suppose $T = 
	\begin{pmatrix}
		1 & \ &\  \\
		& \frac{1}{t} &\  \\
		&   &1 \\  
	\end{pmatrix}$ and $\widetilde{N}={(T \otimes I_3)}^\dag N (T \otimes I_3)$. Then, $\widetilde{N}_{12}=\dfrac{1}{t}N_{12}=\diag(0,0,1)$. So $\widetilde{N}$ has the form in (iii). We replace $M$ with $\widetilde{N}$ and come to the conclusion.

 (iv) By (iii), we assume
	\begin{align}	
		\label{M final}
		&M = 
		\left(\begin{array}{c|c|c}
			\begin{matrix}0 &\  &\  \\  \ &x &\  \\ \ &\  &y \end{matrix}&\ 
			\begin{matrix}0 &\  &\  \\  \ &0 &\  \\ \ &\  &z \end{matrix}&\ 
			\begin{matrix}0 &\  &\  \\  \ &0 &c \\ \ &d &e \end{matrix}\\\hline 
			\begin{matrix}0 &\  &\  \\  \ &0 &\  \\ \ &\  &z \end{matrix}&
			\begin{matrix}u &\  &\  \\  \ &0 &\  \\ \ &\  &v \end{matrix}&\ 
			\begin{matrix}\  &\  &f \\  \ &0 &\  \\ g&\  &h \end{matrix}\\\hline 
			\begin{matrix}0 &\  &\  \\  \ &0 &d^* \\ \ &c^* &e^* \end{matrix}&\ 
			\begin{matrix}\  &\  &g^* \\  \ &0 &\  \\ f^* &\  &h^* \end{matrix}&\ 
			M_{33}\
		\end{array}\right),\\ &\text{where}\ x, y, u, v \in \mathbb{R}_+, z \in \mathbb{R}, c,d,e,f,g,h \in \mathbb{C}.
	\end{align}

	Let $H$ be the matrix obtained by removing the first and fifth rows and columns of $M$:
	\begin{align}	
		\label{N matrix}
		&H=
	\begin{pmatrix}
		A_1 & B & C_1 \\
		B & A_2 & C_2 \\
		{C_1}^\dag  & {C_2}^\dag  & M_{33} \\  
	\end{pmatrix},\notag \\ \text{where}\ 
        A_1 = 
	\begin{pmatrix}
		x & \ \\
		\ & y \\
	\end{pmatrix}, B =& 
	\begin{pmatrix}
		0 & \ \\
		\ & z \\
	\end{pmatrix}, A_2 = 
	\begin{pmatrix}
		u & \ \\
		\ & v \\
	\end{pmatrix}, C_1 = 
	\begin{pmatrix}
		0 & 0 & c \\
		0 & d & e \\
	\end{pmatrix}, C_2 = 
	\begin{pmatrix}
		0 & 0 & f \\
		g & 0 & h \\
	\end{pmatrix}.
	\end{align}
	By $\ine M=(0,2,7)$ in (i), we have $H>0$. So $$H \simeq 
	\begin{pmatrix}
		A_1 & \  & \  \\
		\  & S & \  \\
		\  &\  & M_{33} - C_1^\dag  A_1^{-1} C_1 - (C_2^\dag  - C_1^\dag  A_1^{-1} B) S^{-1} (C_2 - B A_1^{-1} C_1)
	\end{pmatrix},$$
	where $S = A_2-B A_1^{-1} B$. By $H>0$, we have
	\begin{align}
		\label{324 6}
		M_{33} - C_1^\dag  A_1^{-1} C_1 - (C_2^\dag  - C_1^\dag  A_1^{-1} B) S^{-1} (C_2 - B A_1^{-1} C_1) > 0.
	\end{align}
 
Similarly, let $\widetilde{H}$ be the matrix obtained by removing the first and fifth rows and columns of $M^\G$:
\begin{align}\label{324 7}
&\widetilde{H} = 
	\begin{pmatrix}
		A_1 & B & D_1 \\
		B & A_2 & D_2 \\
		{D_1}^\dag  & {D_2}^\dag  & M_{33} \\  
	\end{pmatrix}, \notag \\ \text{where}\ 
        A_1 = 
	\begin{pmatrix}
		x & \ \\
		\ & y \\
	\end{pmatrix}, B =& 
	\begin{pmatrix}
		0 & \ \\
		\ & z \\
	\end{pmatrix}, A_2 = 
	\begin{pmatrix}
		u & \ \\
		\ & v \\
	\end{pmatrix}, D_1 = 
	\begin{pmatrix}
		0 & 0 & d^* \\
		0 & c^* & e^* \\
	\end{pmatrix}, D_2 = 
	\begin{pmatrix}
		0 & 0 & g^* \\
		f^* & 0 & h^* \\
	\end{pmatrix}.
 \end{align}
 Then, $$\widetilde{H} \simeq 
	\begin{pmatrix}
		A_1 & \  & \  \\
		\  & S & \  \\
		\  &\  & M_{33} - D_1^\dag  A_1^{-1} D_1 - (D_2^\dag  - D_1^\dag  A_1^{-1} B) S^{-1} (D_2 - B A_1^{-1} D_1)
\end{pmatrix}.$$
	By $\ine M^\Gamma = (3,2,4)$, we have $\ine \widetilde{H} = (3,0,4)$. We also have $A_1,S>0$. So we have
	\begin{eqnarray}
		\label{neq_2}
		M_{33} - D_1^\dag  A_1^{-1} D_1 - (D_2^\dag  - D_1^\dag  A_1^{-1} B) S^{-1} (D_2 - B A_1^{-1} D_1) < 0.
	\end{eqnarray}

	By \eqref{324 6} and \eqref{neq_2}, we have
	\begin{align}\label{324 8}
		L :=& D_1^\dag  A_1^{-1} D_1 +(D_2^\dag  - D_1^\dag  A_1^{-1} B) S^{-1} (D_2 - B A_1^{-1} D_1)\notag \\ -& C_1^\dag  A_1^{-1} C_1 - (C_2^\dag  - C_1^\dag  A_1^{-1} B) S^{-1} (C_2 - B A_1^{-1} C_1) > 0.
	\end{align}

 By $S>0$, we assume $S^{-1}=\begin{pmatrix}
		s_1 & \ \\
		\ & s_2 \\
	\end{pmatrix}$, where $s_1, s_2 > 0$. We respectively refer to $\epsilon_1,\epsilon_2$ and $\epsilon_3$ as the vectors $(1,0,0)^T,(0,1,0)^T$ and $(0,0,1)^T$.
	By \eqref{N matrix}, \eqref{324 7} and \eqref{324 8}, we have
	$$\epsilon_1^\dag L \epsilon_1 = s_2 (\abs{f}^2 - \abs{g}^2) > 0 \Rightarrow \abs{f} > \abs{g},$$
	$$\epsilon_2^\dag L \epsilon_2= (\frac{1}{y} + \frac{z^2}{y^2} s_2)(\abs{c}^2 - \abs{d}^2) > 0 \Rightarrow \abs{c} > \abs{d}.$$However, 
		$$\epsilon_3^\dag L \epsilon_3 = \frac{1}{x}(\abs{d}^2 - \abs{c}^2) + s_1(\abs{g}^2 - \abs{f}^2) < 0.$$ That is a contradiction with \eqref{324 8}. 
\end{proof}

\begin{lemma}
	\label{lem:324 2}
  Suppose $M=[M_{ij}]\geq 0$ is a $3\times 3$ state and $\ine M^{\Gamma}=(3,2,4)$. If $Me_1 = M^{\Gamma}e_1 = M(e_2+e_6)= M^\Gamma (e_2+e_6) = 0$, then
	\begin{itemize}
	    \item[(i)] $M$ has the form
 \begin{align}\label{sit i}
 M = 
		\left(\begin{array}{c|c|c}
			\begin{matrix}0 &\ &\ \\ \ &x\ &u\ \\ \ &u\ &y\ \end{matrix}&\ 
			\begin{matrix}0 &\ &\ \\ \ &-v &-x \\ \ &-x &-u \end{matrix}&\ 
			\begin{matrix}0 &\ &\ \\ \ &c^*\ &d^*\ \\ \ &d^*\ &b^*\ \end{matrix}\\\hline 
			\begin{matrix}0 &\ &\ \\ \ &-v &-x \\ \ &-x &-u \end{matrix}&
			\begin{matrix}z\ &a\ &\ \\  a^*&w &v \\ \ &v\ &x\ \end{matrix}&\ 
			\begin{matrix}e &f &\ \\  g &h &-c^* \\ \ &-c^* &-d^* \end{matrix}\\\hline 
			\begin{matrix}0 &\ &\ \\  \ &c\ &d\ \\ \ &d\ &b\ \end{matrix}&\ 
			\begin{matrix}e^* &g^* &\ \\ f^* &h^* &-c \\ \ &-c &-d \end{matrix}&\ 
			M_{33} \\
		\end{array}\right),\end{align}
	where $x,y,z,w,u,v \in \mathbb{R}, a,b,c,d,e,f,g,h \in \mathbb{C}$.
	
	\item[(ii)] on the basic of (i), we can make $a=0$.
 \item[(iii)] on the basic of (i) and (ii), we can make $e=b=0$. So $M$ has the form
 \begin{align}\label{sit iii}M = 
		\left(\begin{array}{c|c|c}
			\begin{matrix}0 &\ &\ \\ \ &x\ &u\ \\ \ &u\ &y\ \end{matrix}&\ 
			\begin{matrix}0 &\ &\ \\ \ &-v &-x \\ \ &-x &-u \end{matrix}&\ 
			\begin{matrix}0 &\ &\ \\ \ &c^*\ &d^*\ \\  &d^* & 0 \end{matrix}\\\hline 
			\begin{matrix}0 &\ &\ \\ \ &-v &-x \\ \ &-x &-u \end{matrix}&
			\begin{matrix}z\ &\ &\ \\  &w &v \\ \ &v\ &x\ \end{matrix}&\ 
			\begin{matrix} &f &\ \\  g &h &-c^* \\ \ &-c^* &-d^* \end{matrix}\\\hline 
			\begin{matrix}0 &\ &\ \\  \ &c\ &d\ \\ \ &d\ & 0 \end{matrix}&\ 
			\begin{matrix} &g^* &\ \\ f^* &h^* &-c \\ \ &-c &-d \end{matrix}&\ 
			M_{33} \\
		\end{array}\right),\end{align}
	where $x,y,z,w,u,v \in \mathbb{R}, c,d,f,g,h \in \mathbb{C}$.
   \item[(iv)] there exists a contradiction. So $\ine M^\G \neq (3,2,4).$
 	\end{itemize}
\end{lemma}

\begin{proof}
	(i) By $Me_1=M^\G e_1=0$, $M$ has the form
 $$M=\left(\begin{array}{c|c|c}
		\begin{matrix}0 & & \\  &* &* \\ &* &* \end{matrix}&\ 
		\begin{matrix}0 & & \\  &* &* \\ &* &* \end{matrix}&\ 
		\begin{matrix}0 & & \\  &* &* \\ &* &* \end{matrix}\\\hline 
		\begin{matrix}0 & & \\  &* &* \\ &* &* \end{matrix}&
		\begin{matrix}* &* &* \\ * &* &* \\ *&* &* \end{matrix}&\ 
		\begin{matrix}* &* &* \\ * &* &* \\ *&* &* \end{matrix}\\\hline 
		\begin{matrix}0 & & \\  &* &* \\ &* &* \end{matrix}&\ 
		\begin{matrix}* &* &* \\ * &* &* \\ *&* &* \end{matrix}&\ 
		\begin{matrix}* &* &* \\  *&* &* \\ *&* &* \end{matrix}
	\end{array}\right).$$
 Let $M=(m_{ij})$. We assume $x=m_{22},y=m_{33},z=m_{44},w=m_{55}\in \mathbb{R}$ and $u=m_{23}, v=m_{56},a=m_{45},b=m_{93},c=m_{82},d=m_{92},e=m_{47},f=m_{48},g=m_{57},h=m_{58}\in \mathbb{C}$. So
\begin{align}\label{sit i 1}
M=\left(\begin{array}{c|c|c}
		\begin{matrix}0 & & \\  &x &u \\ &u^* &y \end{matrix}&\ 
		\begin{matrix}0 & & \\  &m_{25} &m_{26} \\ &m_{35} &m_{36} \end{matrix}&\ 
		\begin{matrix}0 & & \\  &c^* &d^* \\ &m_{38} &b^* \end{matrix}\\\hline 
		\begin{matrix}0 & & \\  &m_{52} &m_{53} \\ &m_{62} &m_{63} \end{matrix}&
		\begin{matrix}z &a &m_{46} \\ a^* &w &v \\ m_{64} &v^* &m_{66} \end{matrix}&\ 
		\begin{matrix}e &f &m_{49} \\ g &h &m_{59} \\ m_{67}&m_{68} &m_{69} \end{matrix}\\\hline 
		\begin{matrix}0 & & \\  &c &m_{83} \\ &d &b \end{matrix}&\ 
		\begin{matrix}e^* &g^* &m_{76} \\ f^* &h^* &m_{86} \\ m_{94}&m_{95} &m_{96} \end{matrix}&\ 
		\begin{matrix}m_{77} &m_{78} &m_{79} \\  m_{87}&m_{88} &m_{89} \\ m_{97}&m_{98} &m_{99} \end{matrix}
	\end{array}\right).
 \end{align}
 Since $M(e_2+e_6)=0$, the sum of the second and sixth columns of $M$ is zero. By \eqref{sit i 1}, we have $x+m_{26}=0,u^*+m_{36},m_{46}=0,m_{52}+v=0,m_{62}+m_{66}=0,m_{76}=0,c+m_{86}=0$ and $d+m_{96}=0$. Since $M$ is Hermitian, we have
\begin{align}\label{sit i 2}
M=\left(\begin{array}{c|c|c}
		\begin{matrix}0 & & \\  &x &u \\ &u^* &y \end{matrix}&\ 
		\begin{matrix}0 & & \\  &-v^* &-x \\ &m_{35} &-u^* \end{matrix}&\ 
		\begin{matrix}0 & & \\  &c^* &d^* \\ &m_{38} &b^* \end{matrix}\\\hline 
		\begin{matrix}0 & & \\  &-v &m_{53} \\ &-x &-u \end{matrix}&
		\begin{matrix}z\ &a\ &0 \\ a^*\ &w\ &v \\ 0\ &v^*\ &x \end{matrix}&\ 
		\begin{matrix}e &f &m_{49} \\ g &h &m_{59} \\ 0 &-c^* &-d^* \end{matrix}\\\hline 
		\begin{matrix}0 & & \\  &c &m_{83} \\ &d &b \end{matrix}&\ 
		\begin{matrix}e^* &g^* &0 \\ f^* &h^* &-c \\ m_{94}&m_{95} &-d \end{matrix}&\ 
		\begin{matrix}m_{77} &m_{78} &m_{79} \\  m_{87}&m_{88} &m_{89} \\ m_{97}&m_{98} &m_{99} \end{matrix}
	\end{array}\right).
 \end{align}
By $M^\G(e_2+e_6)=0$, consider the form of $M^\G$ and we have $x+m_{53}=0,u^*-u=0,-v^*+v=0,m_{49}=0,c^*+m_{59}$ and $m_{38}-d^*=0$. So $u,v\in \mathbb{R}$. Substitute $m_{53},m_{49},m_{59},m_{38}$ into \eqref{sit i 2}. Then $M$ satisfies \eqref{sit i}.
 
(ii) If $z=0$, since $M\geq 0$, $\left| \begin{matrix}
    z & a \\
    a^* & w
\end{matrix}\right| =-|a|^2\geq 0$. So $a=0$.

We assume $z\neq 0$. Let $P=\begin{pmatrix}
    1 & -\frac{a}{z} &  \\
      & 1 &  \\
     & & 1   
\end{pmatrix}$ and $N=(I_3 \otimes P)^\dag M(I_3 \otimes P)$. Then 
$$N_{22}=P^\dag M_{22}P=\begin{pmatrix}
    z & 0 & 0\\
    0 & w-\frac{a^*a}{z} & v \\
    0 & v & x
\end{pmatrix},$$ $$N_{23}=P^\dag M_{23}P=\begin{pmatrix}
    e & f-\frac{ae}{z} & 0\\
    g-\frac{a^*e}{z}\ & h-\frac{ag}{z}-\frac{a^*f}{z}+\frac{ea^*a}{z^2} & -c^*\\
    0 & \ -c^* & -d^*
\end{pmatrix}.$$
Moreover, $N_{11}=M_{11},N_{12}=M_{12}$ and $N_{13}=M_{13}$. So $N$ has the form in \eqref{sit i} where $a=0$. We replace $M$ with $N$ and come to the conclusion.

(iii) If $z=0$, since $M\geq 0$, $\left| \begin{matrix}
    z & e\\
    e^* & m_{77}
\end{matrix}\right| =-|e|^2\geq 0$. So $e=0$. Similarly, if $y=0$, then $b=0$. So we assume $z,y\neq 0$. (If one of $z$ and $y$ is zero, we can prove similarly.) 

Let $\phi=-\frac{e}{z}$ and $\lambda=-\frac{eu}{zy}-\frac{b^*}{y}$. We have $\lambda y-\phi u+b^*=0$ and $\phi z+e=0$. Let $Q=\begin{pmatrix}
    1 & & \lambda \\
     & 1 & \phi \\
     & & 1
\end{pmatrix}$ and $N=(Q \otimes I_3)^\dag M(Q \otimes I_3)$. Then 
$$N_{13}=\lambda M_{11}+\phi M_{12}+M_{13}=\begin{pmatrix}
    0 & & \\
     & c^*+\lambda x-\phi v & d^*+\lambda u-\phi x \\
      & d^*+\lambda u-\phi x & 0
\end{pmatrix},$$ $$N_{23}=\lambda M_{12}^*+\phi M_{22}+M_{23}=\begin{pmatrix}
    0 & f+\phi a & 0 \\
    g+\phi a^* & h-\lambda v+\phi w & -c^*-\lambda x+\phi v \\
    0 & -c^*-\lambda x+\phi v & -d^*-\lambda u+\phi x
\end{pmatrix}.$$
Moreover, $N_{11}=M_{11},N_{12}=M_{12},N_{22}=M_{22}$.
We can verify that $N$ has the form in \eqref{sit iii}. Then we replace $M$ with $N$ and come to the conclusion.
 
(iv) Let $H,\widetilde{H}$ be the matrix obtained by removing the first and sixth rows and columns of $M,M^\G$, respectively:	
	\begin{align}	\label{H_matrix_c}
&H=\begin{pmatrix}
		A_1 & B & C_1 \\
		B^\dag & A_2 & C_2 \\
		C_1^\dag & C_2^\dag & M_{33} \\  
	\end{pmatrix} ,\notag \\
 \text{where}\ A_1 = 
	\begin{pmatrix}
		x & u \\
		u & y \\
	\end{pmatrix}, B =& 
	\begin{pmatrix}
		0 & -v \\
		0 & -x \\
	\end{pmatrix}, A_2 = 
	\begin{pmatrix}
		z & \ \\
		\ & w \\
	\end{pmatrix}, C_1 = 
	\begin{pmatrix}
		0 & c^* & d^* \\
		0 & d^* & 0 \\
	\end{pmatrix}, C_2 = 
	\begin{pmatrix}
		0 & f & 0 \\
		g & h & -c^* \\
	\end{pmatrix}.
 \end{align}
 \begin{align} \label{Ha_matrix_c}
 &\widetilde{H}=\begin{pmatrix}
		A_1 & B & D_1 \\
		B^\dag & A_2 & D_2 \\
		D_1^\dag & D_2^\dag & M_{33} \\  
	\end{pmatrix},\notag \\
\text{where}\ A_1 = 
	\begin{pmatrix}
		x & u \\
		u & y \\
	\end{pmatrix}, B =& 
	\begin{pmatrix}
		0 & -v \\
		0 & -x \\
	\end{pmatrix}, A_2 = 
	\begin{pmatrix}
		z & \ \\
		\ & w \\
	\end{pmatrix}, D_1 = 
	\begin{pmatrix}
		0 & c & d \\
		0 & d & 0 \\
	\end{pmatrix}, D_2 = 
	\begin{pmatrix}
		0 & g^* & 0 \\
		f^* & h^* & -c \\
	\end{pmatrix}.
	\end{align}

	By $M\geq 0$ and $\ine M^\G=(3,2,4)$, we have 
 \begin{align}\label{324 13}
 H\geq 0\ \text{and}\ \ine \widetilde{H} = (3,0,4).
 \end{align}
 Because $M_{12}$ is Hermitian in \eqref{sit iii}, by Lemma \ref{basic tran}(iii), we have $\rank(M_{11})=2$. So 
	$A_1=\begin{pmatrix}
		x\ &u\ \\ 
		u\ &y\ \\
	\end{pmatrix}>0$.
 
	For any $\delta>0$, we replace $z, w$ with $z + \delta, w + \delta$ and consider the matrix $\begin{pmatrix}
		A_1 & B \\
		B^\dag & A_2  
	\end{pmatrix}=\begin{pmatrix}
		x & u & 0 & -v \\
            u & y & 0 & -x \\
            0 & 0 & z & 0 \\
            -v & -x & 0 & w
	\end{pmatrix}.$ By $\begin{pmatrix}
		A_1 & B \\
		B^\dag & A_2  
	\end{pmatrix}\geq 0$ and $A_1>0$, all order principal minor determinants of $\begin{pmatrix}
		x & u & 0 & -v \\
            u & y & 0 & -x \\
            0 & 0 & z+\delta & 0 \\
            -v & -x & 0 & w+\delta
	\end{pmatrix}$ are positive. So it is a positive-definite matrix. When $\delta$ are small enough, $\ine \widetilde{H} = (3,0,4)$ is unchanged. So we can assume \begin{align}\label{sit iv 1}\begin{pmatrix}
		A_1 & B \\
		B^\dag & A_2  
	\end{pmatrix}>0. \end{align}
 
Then, we have $$H  \simeq 
	\begin{pmatrix}
		A_1 & \  & \  \\
		\  & S & \  \\
		\  &\  & M_{33} - C_1^\dag A_1^{-1} C_1 - (C_2^\dag - C_1^\dag A_1^{-1} B_1) S^{-1} (C_2 - B_1 A_1^{-1} C_1)
	\end{pmatrix},$$
	
	$$\widetilde{H} \simeq 
	\begin{pmatrix}
		A_1 & \  & \  \\
		\  & S & \  \\
		\  &\  & M_{33} - D_1^\dag A_1^{-1} D_1 - (D_2^\dag - D_1^\dag A_1^{-1} B_1) S^{-1} (D_2 - B_1 A_1^{-1} D_1)
	\end{pmatrix},$$
	where $S = A_2 - B_1^\dag A_1^{-1} B_1$. By \eqref{sit iv 1}, we have $A_1,S>0$. By \eqref{324 13}, we have $$M_{33} - C_1^\dag A_1^{-1} C_1 - (C_2^\dag - C_1^\dag A_1^{-1} B_1) S^{-1} (C_2 - B_1 A_1^{-1} C_1)\geq 0,$$ 
 $$M_{33} - D_1^\dag A_1^{-1} D_1 - (D_2^\dag - D_1^\dag A_1^{-1} B_1) S^{-1} (D_2 - B_1 A_1^{-1} D_1)<0.$$
 
 So 
 \begin{align}\label{324 15}
 L &= D_1^\dag A_1^{-1} D_1  +(D_2^\dag - D_1^\dag A_1^{-1} B_1) S^{-1} (D_2 - B_1 A_1^{-1} D_1) \notag \\ &- C_1^\dag A_1^{-1} C_1- (C_2^\dag - C_1^\dag A_1^{-1} B_1) S^{-1} (C_2 - B_1 A_1^{-1} C_1) > 0.
 \end{align}
	
	Note that $S = A_2 - B_1^\dag A_1^{-1} B_1$ is a diagonal matrix. We suppose $S^{-1} = 
	\begin{pmatrix}
		\lambda & \ \\
		\ & \mu \\
	\end{pmatrix}, A_1^{-1} = 
	\begin{pmatrix}
		\phi & \tau \\
		\tau & \delta \\
	\end{pmatrix}$. Then we compute 
 \begin{align*}
 \bra{2} L \ket{2} &=  (d^*, 0)
	\begin{pmatrix}
		\phi & \tau \\
		\tau & \delta \\
	\end{pmatrix}
	\begin{pmatrix}
		d \\
		0 \\
	\end{pmatrix}+ (0, -c^* + (\phi v + \tau x) d^*)
	\begin{pmatrix}
		\lambda & \ \\
		\  & \mu \\
	\end{pmatrix}
	\begin{pmatrix}
		0 \\
		-c + (\phi v + \tau x)d  \\
	\end{pmatrix}  \\
& - (d, 0)
	\begin{pmatrix}
		\phi & \tau \\
		\tau & \delta \\
	\end{pmatrix}
	\begin{pmatrix}
		d^* \\
		0 \\
	\end{pmatrix} - (0, -c + (\phi v + \tau x) d)
	\begin{pmatrix}
		\lambda & \ \\
		\  & \mu \\
	\end{pmatrix}
	\begin{pmatrix}
		0 \\
		-c^* + (\phi v + \tau x) d^* \\
	\end{pmatrix}\\
 &= 0.
 \end{align*}
 That is a contradiction with \eqref{324 15}.
\end{proof}

By the lemmas, we arrive at the final result of this section.

\begin{theorem}\label{total}
$\mathcal{N}_{3,3}=\{ (1,0,8), (1,1,7), (1,2,6), (1,3,5), (1,4,4), (1,5,3), (2,0,7), $\\
$(2,1,6), (2,2,5), (2,3,4), (3,0,6), (3,1,5), (4,0,5) \}$. 
\end{theorem}
\begin{proof}
   	By Lemma \ref{end trans}--\ref{lem:324 2}, we have $(3,2,4)\notin \mathcal{N}_{3,3}$. By Lemma \ref{le:N_33} and Theorem \ref{thm:414}, we come to the conclusion.
\end{proof}

\section{applications}
\label{sec:app}

In this section, we apply the techniques and results of previous sections to explore some inertia of partial transpose of $12\times12$ positive semidefinite matrices.

Through Lemma \ref{le:(m-1)(n-1)+1}, we study the existence of product vectors in the kernel space. In this way, we can exclude some inertias in $\mathcal{N}_{3,4}$.
\begin{theorem}
    \begin{itemize}
        \item[(i)] $(7,0,5) \notin \mathcal{N}_{3,4}$.
        \item[(ii)] $(7,1,4),(7,2,3) \notin \mathcal{N}_{3,4}$. 
    \item[(iii)] $(8,0,4),(8,1,3),(9,0,3) \notin \mathcal{N}_{3,4}$.
     \end{itemize}
\end{theorem}
\begin{proof}
    (i) We disprove the claim. Assume $M\geq 0$ is a $3\times 4$ state and $\ine M^\G=(7,0,5)$. By Lemma \ref{le:(m-1)(n-1)+1}, the negative space of $M^\G$ has a product vector $\alpha$. By Lemma \ref{le:pro in ker}, we have $M\alpha=0$, which is a contradiction with $\ine M=(7,0,5)$.

    (ii) We disprove the claim. Assume $M\geq 0$ is a $3\times 4$ state and $\ine M^\G=(7,1,4)$ or $(7,2,3)$. Let $N=M+\delta I_9$ where $\delta>0$. When $\delta$ is small enough, we have $N^\G=M^\G+\delta I_9=(7,0,5)$ and $N>0$, which is a contradiction with the nonexistence of $(7,0,5)$ in (i). 

    (iii) Similarly to (i), we can prove $(8,0,4),(9,0,3) \notin \mathcal{N}_{3,4}$. Similarly to (ii), we can prove $(8,1,3) \notin \mathcal{N}_{3,4}$. 
\end{proof}

Through the complete characterization of $\mathcal{N}_{3,3}$ in Theorem \ref{total}, we can also exclude some inertias in $\mathcal{N}_{3,4}$.
\begin{theorem}
    $(6,2,4),(6,3,3),(5,4,3) \notin \mathcal{N}_{3,4}$.
\end{theorem}
\begin{proof}
    Assume $M=[M_{ij}] \geq 0$ is a $3\times 4$ state and $\ine M^\G=(a,b,c)$. Let $P=\diag(0,1,1,1,0,1,1,1,0,1,1,1)$ and $N=P^\dag M P$. $N$ is the submatrix of $M$ by removing the first, fifth and ninth rows and columns. Then by Lemma \ref{product tran}, $N^\G=P^\dag M^\G P$ is the submatrix of $M^\G$. By Lemma \ref{le:projection}, we have  
    \begin{align}\label{thm 1}
    i_+(N^\G)\leq i_+(M^\G)=c.\end{align} 
    For the order-$12$ matrix $M^\G$, by Lemma \ref{sumdif lem}, we have 
    \begin{align*}
    i_-(M^\G)&\leq 
    i_-\left(\begin{array}{c|c|c}
		\begin{matrix}* &* &* &* \\ * &0 & & \\ *& &0 & \\ *& & &0 \end{matrix}&\ 
		\begin{matrix}*&* &* &* \\ &0 & & \\ & &0 & \\ & & &0 \end{matrix}&\ 
		\begin{matrix}*&* &* &* \\ &0 & & \\ & &0 & \\ & & &0 \end{matrix}\\\hline 
		\begin{matrix}*&  & & \\ *&0 & & \\ * & &0 & \\ *& & & 0 \end{matrix}&
		\begin{matrix}0 & & & \\ &0 & & \\ & & 0 & \\ & & & 0 \end{matrix}&\ 
		\begin{matrix}0 & & & \\ &0 & & \\ & & 0 & \\ & & & 0 \end{matrix}\\\hline 
		\begin{matrix}*&  & & \\ *&0 & & \\ * & &0 & \\ *& & & 0\end{matrix}&\ 
		\begin{matrix}0 & & & \\ &0 & & \\ & & 0 & \\ & & & 0 \end{matrix}&\ 
		\begin{matrix}0 & & & \\ &0 & & \\ & & 0 & \\ & & & 0 \end{matrix}
	\end{array}\right)+
 i_-\left(\begin{array}{c|c|c}
		\begin{matrix}0 & & & \\ &0 & & \\ & & 0 & \\ & & & 0 \end{matrix}&\ 
		\begin{matrix}0 & & & \\ * &0 & & \\ * & & 0 & \\ * & & & 0  \end{matrix}&\ 
		\begin{matrix}0 & & & \\ &0 & & \\ & & 0 & \\ & & & 0  \end{matrix}\\\hline 
		\begin{matrix}0 &* &* &* \\ &0 & & \\ & & 0 & \\ & & & 0  \end{matrix}&
		\begin{matrix}* &* &* &* \\ * &0 & & \\ * & & 0 & \\ * & & & 0 \end{matrix}&\ 
		\begin{matrix}* &* &* &* \\ &0 & & \\ & & 0 & \\ & & & 0 \end{matrix}\\\hline 
		\begin{matrix}0 & & & \\ &0 & & \\ & & 0 & \\ & & & 0 \end{matrix}&\ 
		\begin{matrix}* & & & \\ * &0 & & \\ * & & 0 & \\ * & & & 0 \end{matrix}&\ 
		\begin{matrix}0 & & & \\ &0 & & \\ & & 0 & \\ & & & 0 \end{matrix}
	\end{array}\right)\\
&+ i_-\left(\begin{array}{c|c|c}
		\begin{matrix}0 & & & \\ &0 & & \\ & & 0 & \\ & & & 0 \end{matrix}&\ 
		\begin{matrix}0 & & & \\ &0 & & \\ & & 0 & \\ & & & 0  \end{matrix}&\ 
		\begin{matrix}0 & & & \\ * &0 & & \\ * & & 0 & \\ * & & & 0  \end{matrix}\\\hline 
		\begin{matrix}0 & & & \\ &0 & & \\ & & 0 & \\ & & & 0  \end{matrix}&
		\begin{matrix}0 & & & \\ &0 & & \\ & & 0 & \\ & & & 0 \end{matrix}&\ 
		\begin{matrix}0 & & & \\ * &0 & & \\ * & & 0 & \\ * & & & 0 \end{matrix}\\\hline 
		\begin{matrix}0 &* &* &* \\ &0 & & \\ & & 0 & \\ & & & 0 \end{matrix}&\ 
		\begin{matrix}0 &* &* &* \\ &0 & & \\ & & 0 & \\ & & & 0 \end{matrix}&\ 
		\begin{matrix}* &* &* &* \\ * &0 & & \\ * & & 0 & \\ * & & & 0 \end{matrix}
	\end{array}\right)+
 i_-\left(\begin{array}{c|c|c}
		\begin{matrix}0 & & & \\ &* &* &* \\ &* & * &* \\ &* &* & * \end{matrix}&\ 
		\begin{matrix}0 & & & \\ &* &* &* \\ &* & * &* \\ &* &* & * \end{matrix}&\ 
		\begin{matrix}0 & & & \\ &* &* &* \\ &* & * &* \\ &* &* & * \end{matrix}\\\hline 
		\begin{matrix}0 & & & \\ &* &* &* \\ &* & * &* \\ &* &* & *  \end{matrix}&
		\begin{matrix}0 & & & \\ &* &* &* \\ &* & * &* \\ &* &* & * \end{matrix}&\ 
		\begin{matrix}0 & & & \\ &* &* &* \\ &* & * &* \\ &* &* & * \end{matrix}\\\hline 
		\begin{matrix}0 & & & \\ &* &* &* \\ &* & * &* \\ &* &* & * \end{matrix}&\ 
		\begin{matrix}0 & & & \\ &* &* &* \\ &* & * &* \\ &* &* & * \end{matrix}&\ 
		\begin{matrix}0 & & & \\ &* &* &* \\ &* & * &* \\ &* &* & * \end{matrix}
	\end{array}\right)\\
 &\leq 3+i_-(N^\G).
    \end{align*}
    So \begin{align}\label{thm 2}
    i_-(N^\G)\geq a-3.
    \end{align}
    
    (a) If $\ine M^\G=(6,2,4)$, by \eqref{thm 1} and \eqref{thm 2}, we have $i_-(N^\G)\geq 3$ and $i_+(N^\G)\leq 4$. By Theorem \ref{total}, there is no inertia in $\mathcal{N}_{3,3}$ satisfying the conditions. So $(6,2,4)\notin \mathcal{N}_{3,4}$.

        (b) If $\ine M^\G=(6,3,3)$, by \eqref{thm 1} and \eqref{thm 2}, we have $i_-(N^\G)\geq 3$ and $i_+(N^\G)\leq 3$. Similarly to (a), by Theorem \ref{total}, there is no inertia in $\mathcal{N}_{3,3}$ satisfying the conditions. So $(6,3,3)\notin \mathcal{N}_{3,4}$.

            (c) If $\ine M^\G=(5,4,3)$, by \eqref{thm 1} and \eqref{thm 2}, we have $i_-(N^\G)\geq 2$ and $i_+(N^\G)\leq 3$. Similarly to (a), by Theorem \ref{total}, there is no inertia in $\mathcal{N}_{3,3}$ satisfying the conditions. So $(5,4,3)\notin \mathcal{N}_{3,4}$.

\end{proof}

\section{conclusions}
\label{sec:con}

We have shown that the partial transpose of a $9\times9$ semidefinite positive matrix does not have inertia $(3,2,4)$ and $(4,1,4)$. So we have exhausted all inertias of such matrices. We further study the inertia of partial transpose of positive matrices of higher dimensions. Our results help understand the structure of Hermitian matrices and may construct novel entanglement witnesses in quantum information science.

\section*{Acknowledgments}

Authors were supported by the NNSF of China (Grant No. 11871089).

\bibliographystyle{unsrt}

\bibliography{changchun=inertia}

\begin{thebibliography}{10}

\bibitem{Micha1998Mixed}
Micha, Horodecki, Pawe, Horodecki, Ryszard, and Horodecki.
\newblock Mixed-state entanglement and distillation: Is there a "bound"
  entanglement in nature?
\newblock {\em Physical Review Letters}, 80(24):5239–5242, 1998.

\bibitem{GUHNE20091}
Otfried Gühne and Géza Tóth.
\newblock Entanglement detection.
\newblock {\em Physics Reports}, 474(1):1--75, 2009.

\bibitem{2007Quantum}
Ryszard Horodecki, Pawel Horodecki, Michal Horodecki, and Karol Horodecki.
\newblock Quantum entanglement.
\newblock {\em Review of Modern Physics}, 2007.

\bibitem{2013Negative}
S.~Rana.
\newblock Negative eigenvalues of partial transposition of arbitrary bipartite
  states.
\newblock {\em PHYSICAL REVIEW A}, 87(5):1--4, 2013.

\bibitem{2013Non}
N.~Johnston.
\newblock Non-positive-partial-transpose subspaces can be as large as any
  entangled subspace.
\newblock {\em Physical Review A}, 87(6):459--496, 2013.

\bibitem{Peres1996}
A.~Peres.
\newblock Separability criterion for density matrices.
\newblock {\em Phys. Rev. Lett.}, 77:1413, 1996.

\bibitem{2012Qubit}
C.~Lin and Dragomir~Z. Djokovic.
\newblock Qubit-qudit states with positive partial transpose.
\newblock {\em Physical Review A}, 86(6):12510--12517, 2012.

\bibitem{horodecki1997}
P.~Horodecki.
\newblock Separability criterion and inseparable mixed states with positive
  partial transposition.
\newblock {\em Phys. Lett. A}, 232:333, 1997.

\bibitem{2000Entanglement}
B.~M. Terhal and Kgh Vollbrecht.
\newblock Entanglement of formation for isotropic states.
\newblock {\em Physical Review Letters}, 85(12):2625, 2000.

\bibitem{2019Design}
D.~Amaro and M~Müller.
\newblock Design and experimental performance of local entanglement witness
  operators.
\newblock 2019.

\bibitem{2020Measurement}
Zheng~Da Li, Qi~Zhao, Rui Zhang, Li~Zheng Liu, and Jian~Wei Pan.
\newblock Measurement-device-independent entanglement witness of tripartite
  entangled states and its applications.
\newblock {\em Physical Review Letters}, 124(16), 2020.

\bibitem{2002Computable}
G.~Vidal and R.~F. Werner.
\newblock Computable measure of entanglement.
\newblock {\em Physical Review A}, 65(3):032314, 2002.

\bibitem{2008Universal}
R.~Augusiak, M.~Demianowicz, and P.~Horodecki.
\newblock Universal observable detecting all two-qubit entanglement and
  determinant-based separability tests.
\newblock {\em Physical Review A}, 2008.

\bibitem{2020Inertias}
Y.~Shen, L.~Chen, and L.~J. Zhao.
\newblock Inertias of entanglement witnesses.
\newblock {\em Journal of Physics A: Mathematical and Theoretical},
  53(48):485302 (27pp), 2020.

\bibitem{3x3inertia2022changchun}
Changchun Feng, Lin Chen, Chang Xu, and Yi~Shen.
\newblock Inertia of two-qutrit entanglement witnesses.
\newblock {\em Linear and Multilinear Algebra}, 0(0):1--23, 2022.

\bibitem{Chen2013}
Chen L and Djokovic~D Z.
\newblock Properties and construction of extreme bipartite states having
  positive partial transpose, 2013.

\bibitem{David2003Inertia}
David, A, Gregory, , , Brenda, Heyink, , , Kevin, N.~Vander, and Meulen.
\newblock Inertia and biclique decompositions of joins of graphs.
\newblock {\em Journal of Combinatorial Theory}, 2003.

\end{thebibliography}

\end{document}